\documentclass[printer]{aa}

\usepackage{mhchem}
\usepackage{bm}
\usepackage{siunitx}
\usepackage{tikz, mathtools}
\usepackage{tikz-3dplot} 
\usepackage{multirow, amsmath}
\usetikzlibrary{shapes.geometric, arrows,arrows.meta,positioning,quotes,automata,decorations.text}
\usepackage{array}
\usepackage{tabu}
\usepackage{tabularx}
\usepackage{makecell}

\usetikzlibrary{plotmarks}

\usepackage[varg]{txfonts}
\usepackage{natbib,twoopt}
\usepackage[breaklinks=true]{hyperref} 
\bibpunct{(}{)}{;}{a}{}{,}             
\makeatletter
  \newcommandtwoopt{\citeads}[3][][]{\href{http://adsabs.harvard.edu/abs/#3}%
    {\def\hyper@linkstart##1##2{}%
     \let\hyper@linkend\@empty\citealp[#1][#2]{#3}}}
  \newcommandtwoopt{\citepads}[3][][]{\href{http://adsabs.harvard.edu/abs/#3}%
    {\def\hyper@linkstart##1##2{}%
     \let\hyper@linkend\@empty\citep[#1][#2]{#3}}}
  \newcommandtwoopt{\citetads}[3][][]{\href{http://adsabs.harvard.edu/abs/#3}%
    {\def\hyper@linkstart##1##2{}%
     \let\hyper@linkend\@empty\citet[#1][#2]{#3}}}
  \newcommandtwoopt{\citeyearads}[3][][]%
    {\href{http://adsabs.harvard.edu/abs/#3}
    {\def\hyper@linkstart##1##2{}%
     \let\hyper@linkend\@empty\citeyear[#1][#2]{#3}}}
\makeatother

\definecolor{edits}{rgb}{0,0,0}
\definecolor{edits2}{rgb}{0,0,0}
\definecolor{comments}{rgb}{0,0,0}
\definecolor{delete1}{rgb}{0,0,0}

\begin{document}
 \title{Multi-instrument analysis of far-ultraviolet aurora in the southern hemisphere of comet 67P/Churyumov-Gerasimenko}
 
  \author{P. Stephenson \inst{1}
\and M. Galand \inst{1}
\and P. D. Feldman \inst{2}
\and A. Beth \inst{3}
\and M. Rubin \inst{4}
\and D. Bockel\'ee-Morvan \inst{5}
\and N. Biver \inst{5}
\and Y.-C Cheng \inst{5}
\and J. Parker \inst{6}
\and J. Burch \inst{7}
\and F. L. Johansson \inst{8}
\and A. Eriksson \inst{8}
}

 \institute{Department of Physics, Imperial College London, London, SW7 2AZ, UK
 \and Physics and Astronomy, The Johns Hopkins University, 3400 N. Charles Street, Baltimore, MD 21218, USA
 \and Department of Physics, Ume{\aa} University, 901 87 Ume{\aa}, Sweden
 \and Physikalisches Institut, University of Bern, Sidlerstrasse 5, 3012 Bern, Switzerland
 \and LESIA, Observatoire de Paris, Université PSL, CNRS, Sorbonne Universit\'e, Universit\'e de Paris, 5 place Jules Janssen, 92195 Meudon, France
 \and Southwest Research Institute, Department of Space Studies, Suite 300, 1050 Walnut Street, Boulder, CO 80302, USA
 \and SouthWest Research Institute, P.O. Drawer 28510, San Antonio, TX 78228-0510, USA
 \and Swedish Institute of Space Physics, {\AA}ngstr\"om Laboratory, L\"agerhyddsv\"agen1, 75237 Uppsala,
20 Sweden
 }
 
 \titlerunning{Multi-Instrument Analysis of FUV Aurora in the Southern Hemisphere of Comet 67P/C-G}
\date{Received XXXX / Accepted XXXX}

\abstract{}{We aim to determine whether dissociative excitation of cometary neutrals by electron impact is the major source of far-ultraviolet (FUV) emissions at comet 67P/Churyumov-Gerasimenko in the southern hemisphere at large heliocentric distances, both during quiet conditions and impacts of corotating interaction regions observed in the summer of 2016.}{\textcolor{comments}{{We combined multiple datasets from the Rosetta mission through a multi-instrument analysis}} to complete the first forward modelling of FUV emissions in the southern hemisphere of comet 67P and compared modelled brightnesses to observations with the Alice FUV imaging spectrograph. We modelled the brightness of OI1356, OI1304, Lyman-$\beta$, CI1657, and CII1335 emissions, which are associated with the dissociation products of the four major neutral species in the coma: \ce{CO2}, \ce{H2O}, \ce{CO}, and \ce{O2}. The suprathermal electron population was probed by the Ion and Electron Sensor of the Rosetta Plasma Consortium (RPC/IES) and the neutral column density was constrained by several instruments: the Rosetta Orbiter Spectrometer for Ion and Neutral Analysis (ROSINA), the Microwave Instrument for the Rosetta Orbiter (MIRO) and the Visual InfraRed Thermal Imaging Spectrometer (VIRTIS).}{The modelled and observed brightnesses of the FUV emission lines agree closely when viewing nadir and dissociative excitation by electron impact is shown to be the dominant source of emissions away from perihelion. The CII1335 emissions are shown to be consistent with the volume mixing ratio of CO derived from ROSINA. When viewing the limb during the impacts of corotating interaction regions, the model reproduces brightnesses of OI1356 and CI1657 well, but resonance scattering in the extended coma may contribute significantly to the observed Lyman-$\beta$ and OI1304 emissions. The correlation between variations in the suprathermal electron flux and the observed FUV line brightnesses when viewing the comet’s limb suggests electrons are accelerated on large scales and that they originate in the solar wind. This means that the FUV emissions are auroral in nature.}{}


\maketitle

\section{Introduction}\label{sec: Intro}
Auroras, most familiarly observed at high latitudes over the northern and southern regions of Earth, have been detected at \textcolor{comments}{{several}} bodies in the Solar System. Auroral emissions are generated by (usually charged) extra-atmospheric particles colliding with an atmosphere, causing excitation \citep{Galand2002}.  At Earth, other magnetised planets, and the Jovian moon Ganymede, the magnetospheric structure restricts entry of these extra-atmospheric particles into the atmosphere, confining auroras to regions with open field lines. However, comets are unmagnetised \citep{Heinisch2019}, so they exhibit more similarities to regions of Mars with no crustal magnetisation, where diffuse auroras have been seen \citep{Schneider2015}. \newline
The Rosetta mission \citep{Glassmeier2007} observed comet 67P/Churyumov-Gerasimenko (hereafter 67P) from within the coma throughout the two-year escort phase, allowing measurement of cometary emissions from a new, close perspective. Earth-based observations of comets in the far-ultraviolet (FUV) with the Hubble Space Telescope \citep[HST;][]{Lupu2007, Weaver2011} and the Far Ultraviolet Spectroscopic Explorer \citep[FUSE;][]{Feldman2002, Weaver2002} have not seen evidence of aurora at comets. The analysis of FUV emission spectra from HST and FUSE indicated that photodissociation and resonance scattering were the dominant sources of emissions at these wavelengths. However, these Earth-based observations are limited to active comets, with an outgassing rate of $Q>10^{28}$~\si{\per\second}, which are close to the Sun ($<2$~\si{\astronomicalunit}). Rosetta provided an opportunity to observe a comet further from the Sun ($>3$ \si{\astronomicalunit}) and at much lower levels of activity  ($Q<10^{26}$~\si{\per\second}) than was previously possible \citep{Laeuter2018}.  \newline
The Alice FUV imaging spectrograph \citep{Stern2007} onboard Rosetta was used to measure emission spectra from within the coma of 67P, which contrasted with the Earth-based measurements of cometary emissions. An analysis of \textcolor{comments}{{the emission line ratios}} in FUV spectra has suggested that the dissociative excitation of cometary neutrals by electron impact (\ce{e + X} for a cometary molecule \textcolor{edits}{-} X) was a significant source of FUV emissions \citep{Feldman2015}. Dissociative excitation is driven by suprathermal electrons which have energies greater than the high threshold energies ($>14$ \si{\electronvolt}) for these processes \citep{McConkey2008, Ajello1971, Mumma1972}. Suprathermal electrons have been observed in the coma of 67P throughout the escort phase with the electron spectrometer  \citep{Burch2007} and do not follow a Maxwellian distribution \citep{Clark2015,Broiles2016a}. There is also a thermal population of cold electrons ($<1$ \si{\electronvolt}) which have been observed throughout the escort phase \citep{Eriksson2017, Gilet2020} but their energy is too low to be able to contribute to the FUV emissions. \newline
\textcolor{edits}{Early FUV spectra taken in the northern hemisphere summer were consistent with the impact of electrons on water \citep{Feldman2018}. The outgassing of \ce{H2O} is closely linked to the illumination conditions of the nucleus and exhibits seasonal trends in its production rate \citep{Fink2016, Laeuter2018}. Pre-perihelion, in the northern hemisphere summer, \ce{H2O} is the dominant outgassed neutral species \citep{Hassig2015, Laeuter2018, Bockelee-Morvan2016}, whilst there is also a significant presence of \ce{O2} \citep{Bieler2015}.\newline
Unlike over the northern hemisphere, the post-perihelion spectra analysed over the southern hemisphere summer exhibit strong carbon lines and hence they are driven mostly by electron impact on \ce{CO2}. This reflects the hemispherical asymmetry in the composition of 67P's coma that has been observed at large heliocentric distances with the mass spectrometer \citep{Balsiger2007} as well as with the infrared \citep{Coradini2007} and sub-mm \citep{Gulkis2007} spectrometers onboard Rosetta. Throughout the mission, the outgassing of \ce{CO2} and, to a lesser extent, \ce{CO} were larger in the southern hemisphere than in the northern hemisphere \citep{Laeuter2018}, which reflects an inhomogeneity in the surface of the nucleus. The significant increase in the outgassing of both \ce{CO2} and \ce{CO} post-perihelion \citep{Gasc2017MNRAS, Biver2019} may also result from the exposure of a more pristine surface layer of the nucleus, due to erosion around perihelion \citep{Fink2016, Filacchione2016}.} \newline
 \cite{Chaufray2017} showed that HI-Ly-$\beta$ emissions observed by the \textcolor{edits}{Alice} FUV spectrograph exhibit some correlation with remote measurements of the water column density from \textcolor{edits}{the IR spectrometer on Rosetta}. This demonstrated the dependence of FUV emission brightness on the column density of water along the line of sight. They also calculated the Ly-$\beta$ brightness, assuming a Maxwellian distribution of suprathermal electrons with a constant temperature (17 \si{\electronvolt}) and density (20 \si{\per\cubic\centi\metre}). However, this does not account for large variations (by a factor of 100) that have been observed in the suprathermal electron flux or the non-thermal distribution of electrons. \cite{Chaufray2017} also found that away from perihelion the suprathermal electron flux does not seem to vary with cometocentric distance, in contrast to the total electron density which varies approximately with $1/r$ \citep{Galand2016, Heritier2017Vert}.\newline
 \textcolor{edits}{\cite{RaghuramBhardwaj2020} have analysed the brightness of FUV emissions using a photochemical model, with application to a high outgassing regime ($Q>10^{27}$ s$^{-1}$). However, they model significant emissions from several photodissociation processes that require spin-forbidden transitions, and therefore do not occur. When the analysis is applied to a larger heliocentric distance (1.99~\si{\astronomicalunit}), they cannot explain the observed FUV emission brightnesses.} \newline
\cite{Galand2020} employed a multi-instrument analysis to combine FUV brightnesses with in situ or remote neutral gas observations and in situ measurements of the suprathermal electron flux. \textcolor{edits}{This work focused on the northern, summer hemisphere of comet 67P at large helicoentric distances where \ce{H2O} is the major neutral species, demonstrating that electron impact on \ce{H2O} and \ce{O2} are the dominant sources of emissions.} \cite{Galand2020} concluded that emissions are driven by electrons which have been accelerated on large scales rather than locally heated. Acceleration of solar wind electrons by an ambipolar field is the most likely candidate \citep{Deca2017, Deca2019}, meaning these emissions are auroras, a phenomenon which had not previously been observed at a comet in the FUV. \newline
In the \textcolor{edits}{non-illuminated} southern hemisphere, the FUV emission spectra are very different to those in the northern hemisphere, with much stronger emissions of atomic carbon lines and molecular bands of \ce{CO} \citep{Feldman2018}. There has been no forward modelling to determine whether the FUV emissions in the southern hemisphere are also driven by dissociative excitation of cometary neutrals or to understand which neutral species are key to each emission line. We propose to apply an extension of the multi-instrument analysis of \cite{Galand2020} to model the FUV emissions in the southern hemisphere of comet 67P at large heliocentric distances. \newline 
The brightest emission from the coma is Ly-$\alpha$ at 1216 \si{\angstrom} but due to the high flux of Ly-$\alpha$ photons during the mission, the detector degraded significantly at this wavelength throughout the mission. We model the brightness of emissions in the strongest remaining atomic emission lines. These are associated with atomic transitions of oxygen (OI1356, OI1304), hydrogen (Ly-$\beta$), and carbon (CI1657 and CII1335), which are the dissociation products of the four major neutral species in the coma \citep[\ce{CO2}, \ce{H2O}, \ce{CO}, \ce{O2};][]{Gasc2017MNRAS,Laeuter2018}.\newline
Throughout the duration of Rosetta's escort phase, many solar events, such as Coronal Mass Ejections (CMEs) and Corotating Interaction Regions (CIRs) reached 67P, generating enhancements in the suprathermal electron population in the coma \citep{Edberg2016a, Witasse2017, Hajra2018, Goetz2019}. The variation of emissions during these events has been observed by \cite{Feldman2015} and \cite{Noonan2018}, although the solar event which \cite{Feldman2015} analysed was not identified until after publication \citep{Witasse2017}. \textcolor{edits}{Both of these events were observed in the northern hemisphere when \ce{H2O} was the dominant outgassing species, with the CIR of \cite{Feldman2015} occurring early in the mission and the CME of \cite{Noonan2018} arriving when 67P was close to perihelion.} \cite{Noonan2018} qualitatively compared the 'warm' (\textcolor{comments}{{5-100 \si{\electronvolt}}}) electron density to the brightness of several atomic FUV lines (OI1356, OI1304, Ly-$\beta$ and CI1657) during the arrival of a CME near perihelion. \newline
In the present study, we focus on CIRs observed at 67P throughout the summer of 2016 \citep{Hajra2018}. These are formed when a fast solar wind stream interacts with the slow solar wind that precedes it, generating a region of compression. The compression causes an increase in the electron number density, while shock structures can lead to further heating of electrons at large heliocentric distances \citep{Smith1976}. The CIR is seen periodically ($ \sim\! 25$\si{\day}) from June to September 2016, because of its solar corotation. Electron impact is the dominant ionisation process during these CIRs and contributes significantly to the total plasma density \citep{Heritier2018a}, but there has not been a quantitative assessment of their impact on FUV emissions so far.\newline
In this paper, we utilise a multi-instrument analysis to model the brightness of FUV emission lines in the southern hemisphere of comet 67P at large heliocentric distances, with direct comparison to observed brightnesses from the FUV spectrograph onboard Rosetta. In Section \ref{sec: Methods}, we introduce the methodology used in the study to calculate the brightness of each emission line, using measurements of neutral gas composition and density as well as measurements of the suprathermal electron flux. In Section \ref{sec: Nadir}, we apply this analysis in the southern hemisphere during quiet periods, outside of solar events, both pre- and post-perihelion. We then model FUV emissions from the coma during the August and July occurrences of the CIR in the summer of 2016 in Section \ref{sec: CIRs}. In Section \ref{sec: Conclusion}, we compare our results with those obtained in the northern hemisphere and discuss the implications of our findings on our understanding of the cometary environment and on future analysis of cometary FUV emissions.
\section{Methods}\label{sec: Methods}
\subsection{Multi-instrument analysis} \label{sec: multi int}
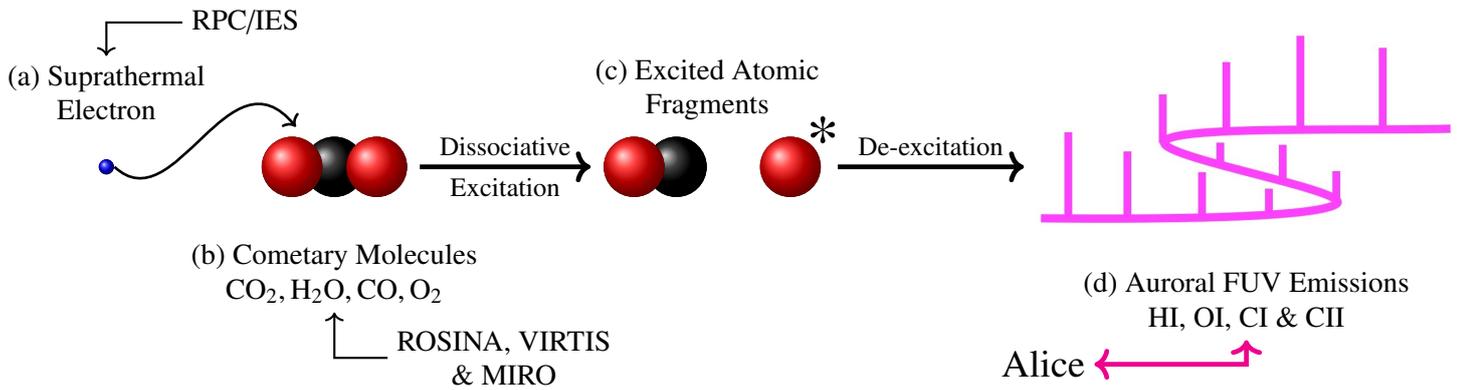
\begin{figure*}[htbp]
    \centering
\begin{tikzpicture}[node distance= 5cm,
arrow/.style = {-{Triangle[]},thick},baseline=(current  bounding  box.center)]
\tikzstyle{species} = [rectangle, minimum width=1cm, minimum height=1cm, text centered]
\tikzstyle{electron} = [circle, fill, ball color = blue, inner sep=0pt, minimum size=0.2cm] 
\tikzstyle{carbon} = [circle, fill, ball color = black, inner sep=0pt,minimum size=0.8cm] 
\tikzstyle{oxygen} = [circle, fill, ball color = red, inner sep=0pt, minimum size=0.8cm] 


\node [electron] (hot_electron) at (0,4) {}; 
\node [carbon] (molecule) at (3,4) {};
\node [oxygen, left = 0.15 of molecule.center] (mol_pt_2a) {};
\node [oxygen, right = 0.15 of molecule.center] (mol_pt_2b) {};

\node [above right = 2 of hot_electron.north east] (c2) {};
\node [below right = 1 of hot_electron.north east] (c1) {};
\node [above left = 0.4 of molecule.center] (c3) {};
\draw [->, line width = 1pt] (hot_electron.east) ..  controls (c1) and (c2) .. (c3);
\node [carbon] (fragment1a) at (7.5,4) {};
\node [oxygen, left = 0.15 of fragment1a.center] (fragment1b) {};
\node[oxygen] (fragment2) at (9,4) {};

\node [inner sep = 0pt] (FUV) at (15,4.5) {\includegraphics[width = 0.3\textwidth ]{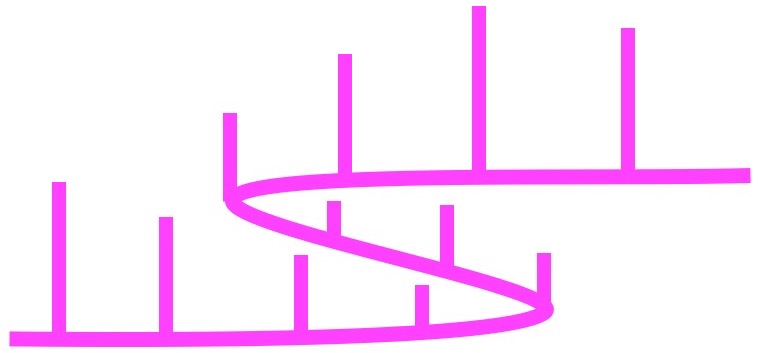}};
\node [right = 1 of molecule.base] (arrow1start) {};
\node [right = 2 of arrow1start] (arrow1end) {};
\draw [->, line width = 1.8] (arrow1start.base) -- node[pos = 0.5, above]{ Dissociative} node[pos = 0.5, below]{Excitation} (arrow1end.base);

\node [right = 0.5 of fragment2.base] (arrow2start) {};
\node [right = 2.2 of arrow2start] (arrow2end) {};
\draw [->, line width = 1.8] (arrow2start.base) -- node[pos = 0.5, above] {De-excitation} (arrow2end.base);
base);

\node [rectangle, above = 0.5 of hot_electron.base, font = \large, align =center] (SupElec) {(a) Suprathermal \\ Electron};
\node [rectangle, below = 0.5 of molecule, font = \large, align = center] (ComMol) {(b) Cometary Molecules \\ \ce{CO2, H2O, CO, O2}};

\node [rectangle, above = 0.5 of fragment1a.east, font = \large, align =center] (ExFragments) {(c) Excited Atomic\\ Fragments};
\node [rectangle, below = 0.5 of FUV.base, font = \large, align = center] (Auroral) {(d) Auroral FUV Emissions \\ HI, OI, CI \& CII};

\node [above right = 1 and 1 of SupElec.base, font = \large, align = center] (IES) {RPC/IES};
\node [below right = 0.3 and 0.7 of ComMol.base, font = \large, align = center] (ColDens) {ROSINA, VIRTIS \\ \& MIRO};
\draw [->, thick] (IES) -| (SupElec);
\draw [->, thick] (ColDens) -| (ComMol);
\node [below left = 0.2 and 2 of Auroral.base, font = \Large] (Alice) {Alice};
\draw [<->, line width = 1.8pt, magenta] (Alice)-| (Auroral);
\node [rectangle, inner sep = 0mm, font = \Huge, above right = 0.48 and 0.2 of fragment2.south] (Exc){*};
\end{tikzpicture}
    \caption{Schematic of the multi-instrument analysis used in this study to model FUV emissions driven by electron impact on cometary neutrals. From left to right: (a) Suprathermal electrons present within the coma were measured using RPC/IES (see Section \ref{sec: electron flux}). (b) Neutral gas molecules in the coma. There were four major neutral species seen at 67P: \ce{CO2, H2O, CO, and O2} (see Section \ref{sec: Neutral Column}). (c) A collision between a suprathermal electron and a cometary molecule causes the molecule to dissociate. A neutral fragment and an excited atom are produced. (d) The excited atom de-excites, releasing a photon in the FUV. These photons were observed with the Alice FUV imaging spectrograph (see Section \ref{sec: Alice Obs}).}
    \label{fig: multi int}
\end{figure*}

In this study, we employed an extension of a multi-instrument analysis developed by \cite{Galand2020} to model the emissions driven by dissociative excitation of cometary neutrals by electron impact. The analysis brings together distinct datasets from several instruments onboard Rosetta. The process of dissociative excitation by electron impact is outlined in Fig.~\ref{fig: multi int}, along with a qualitative description of how each instrument contributes to the analysis.

The equation underlying the methodology allows direct comparison of the brightness derived from the observations, by the Alice FUV imaging spectrograph (see Section \ref{sec: Alice Obs}), with the brightness, $B^X$ \textcolor{comments}{{[R, 1 rayleigh $= 10^6/4\pi$ photons cm$^{-2}$s$^{-1}$sr$^{-1}$]}}, of the atomic emission line, X (OI, CI, CII, HI), calculated as follows:
\begin{equation}\label{eq: model brightness}
B^X =10^{-6}\sum\limits_l N_{l} \int\limits_{E_{Th,l}}^{E_{Max}}\! \sigma^X_l(E) J(E) \, \mathrm{d}E =10^{-6}\sum\limits_l N_l \nu_l^X
\end{equation}
where $N_l$ \textcolor{comments}{{[cm$^{-2}$]}} is the column density of each neutral species, $l$ (\ce{CO2}, \ce{H2O}, \ce{CO} and \ce{O2}), along the line of sight of the FUV spectrograph and the summation is over each of the major species found in the coma of 67P (see Section \ref{sec: Neutral Column}). \textcolor{comments}{{Equation \ref{eq: model brightness} is predicated on the assumption that the suprathermal electron particle flux, $J(E)$ [cm$^{-2}$s$^{-1}$eV$^{-1}$], is constant throughout the column in question (see Section \ref{sec: electron flux})}}. The emission frequency, $\nu_l^X$ \textcolor{comments}{{[s$^{-1}$]}}, is derived from the emission cross-section, $\sigma^X_l(E)$ \textcolor{comments}{{[cm$^2$]}}, for each emission line, $X$, and neutral species, $l$, and from the suprathermal electron particle flux (see Section \ref{sec: emission freq}).

\subsection{Observed brightness of FUV emission lines} \label{sec: Alice Obs}

\begin{figure}[htbp]
\begin{tikzpicture}
\node [inner sep = 0pt] (Spectrum) at (0,0) {\includegraphics[width = 0.5\textwidth]{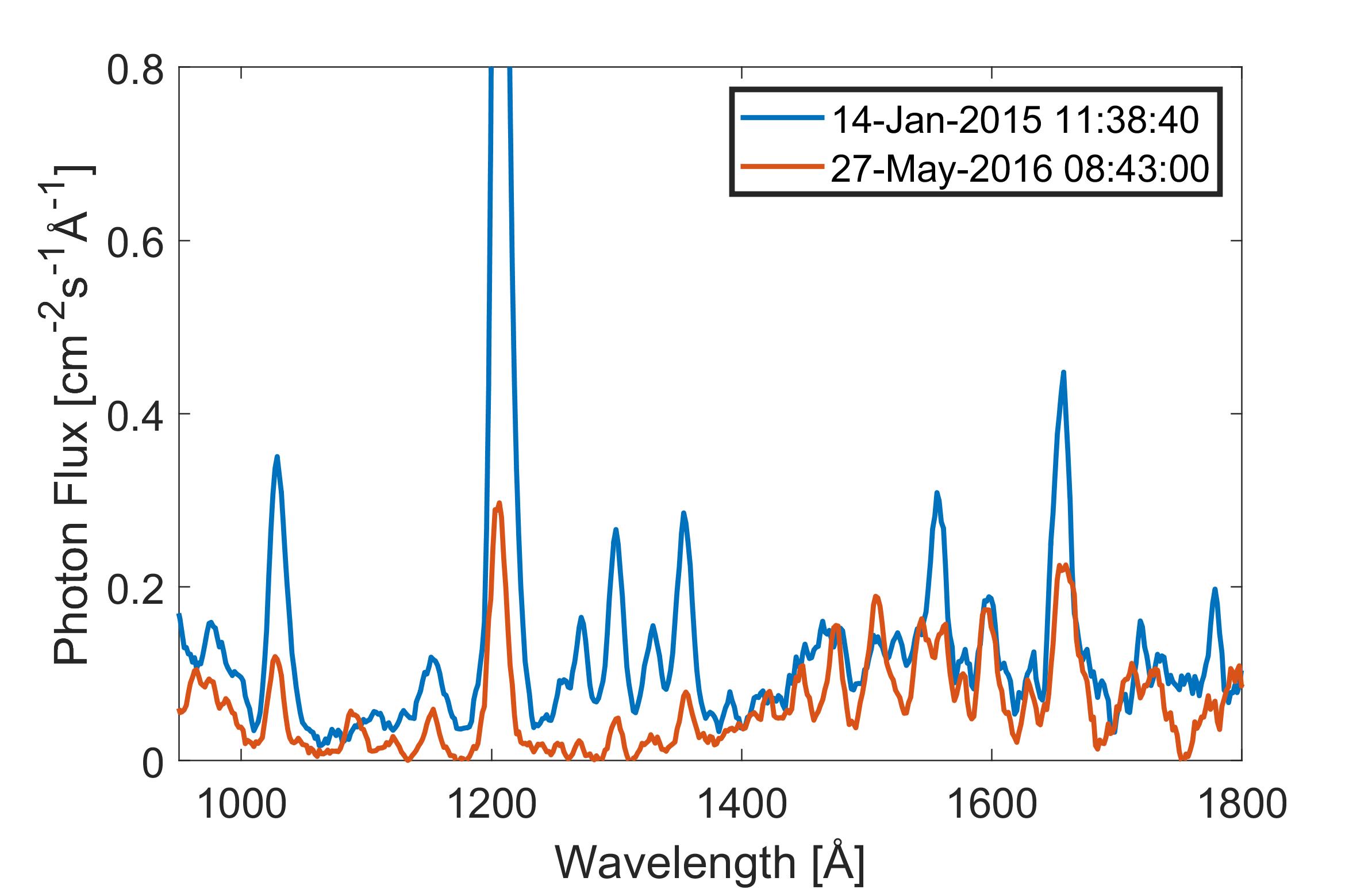}};
\node (LyBbot) at (-2.75,0) {};
\node [above = 0.5 of LyBbot, fill=white] (LyBtop) {Ly-$\beta$};

\node (OI1304bot) at (-0.47,-0.35) {};
\node [above = 0.5 of OI1304bot] (OI1304top)  {};

\node (OI1356bot) at (-0.02,-0.35) {};
\node [above = 0.5 of OI1356bot] (OI1356top) {};

\node (CIIbot) at (-0.23,-1) {};
\node [above = 1.6 of CIIbot] (CIItop) {CII1335};

\node (CIbot) at (2.54,0.6) {};
\node [above = 0.3 of CIbot] (CItop) {CI1657};

\draw [black, thick] (LyBbot) -- (LyBtop);
\draw [black, thick] (OI1304bot) -- (OI1304top);
\draw [black, thick] (OI1356bot) -- (OI1356top);
\draw [black, thick] (CIIbot) -- (CIItop);
\draw [black, thick] (CIbot) -- (CItop);

\node [above left = 0.01 of OI1304top.south east, fill=white, fill opacity=0.8, text opacity=1] (OI1304) {OI1304};
\node [above right = 0.01 of OI1356top.south west] (OI1356) {OI1356};

\end{tikzpicture}
\caption{FUV spectrum measured by Alice during two nadir scans during quiet periods. The spectra \textcolor{comments}{{have been coadded over four rows in the Alice slit}} and smoothed with a five point moving average to minimise the noise in the spectra. The key emissions selected in this study have been highlighted.}
\label{fig: alice spectra}
\end{figure}
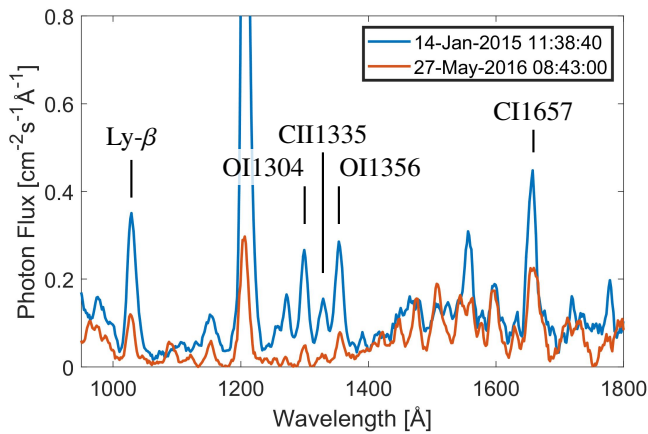

The Alice FUV imaging spectrograph \citep{Stern2007} observed emissions in the range 700 {\AA} - 2050 {\AA}, with a spectral resolution of 8 {\AA} in the slit centre. The slit comprised 32 rows (0 to 31) with Row 15 at the centre, each row subtending $0.3^{\circ}$ along the slit axis. The slit had a dogbone-like shape as the central rows (13 to 17) have a width of $0.05^{\circ}$, whereas the outer rows \textcolor{edits}{($\ge 19$ and $\le 12$)} had a width of $0.10^{\circ}$ \citep{Feldman2015}. Rows 12 and 18 were transitional between these two widths, hence they have been avoided in the present analysis. Typically, the spectrograph scans lasted either approximately five or ten minutes, but to improve the signal-to-noise ratio (S/N), several consecutive spectra can be co-added. \newline
The emission spectra exhibit an odd-even oscillation between rows \citep{Chaufray2017}, so even numbers of rows were co-added to minimise the impact of this aberration. 
When using a nadir viewing, during quiet periods, 67P was fairly stationary in the Alice field-of-view over several scans. As such, the co-added scans, ranging from 20 to 100 \si{\minute}, probed a small region of the coma within each viewing period. \newline
In the cases of solar events, the temporal variation in the brightness of each emission line is highly relevant. Therefore, each spectrum retrieved from Alice has been considered individually, whilst several adjacent rows were combined. To capture some of the spatial variability in the emissions, the brightness was evaluated in three different regions of the Alice viewing slit. \newline
In the present study, we considered emissions from five multiplets (as illustrated in Fig.~\ref{fig: alice spectra}): Lyman-$\beta$, OI1304, OI1356, CI1657 and CII1335. We have not considered emissions \textcolor{edits}{of} Lyman-$\alpha$ as the contribution to this line from the \textcolor{edits}{interplanetary medium} (IPM) is very strong, and the detector of the FUV spectrograph degraded significantly at this wavelength throughout the mission. The contribution of the IPM to the Lyman-$\beta$ brightness is of the order of 2\,R when looking off-limb, whereas the strength of the IPM Ly-$\alpha$ is 300 times larger \citep{Feldman2015}. \newline
There are also other emission features that overlap with the lines of interest and which, therefore, must be considered. The oxygen line at 1027 \si{\angstrom} is not well resolved from the emissions of Ly-$\beta$. The CO Fourth Positive Group (4PG) emits in the range $1400 - 1800${\AA}  and has several bands which can contribute to the emissions observed at 1657{\AA}. Those bands which emit significantly in this range are (3,4) at 1648{\AA}, (0,2) at 1653{\AA} and (1,3) at 1670{\AA}  \citep{Beegle1999}. 

\subsection{Neutral column density} \label{sec: Neutral Column}
The column density of the neutral gas along the line-of-sight of the FUV spectrograph can be calculated with several different methods depending on the viewing geometry of Rosetta. When observing nadir, we can extrapolate from in-situ total neutral density measurements by the Rosetta Orbiter Spectrometer for Ion and Neutral Analysis  \citep[\textcolor{edits}{ROSINA,}][]{Balsiger2007}. We derived the neutral composition from measurements by the Double Focusing Mass Spectrometer (ROSINA/DFMS) but this was not available at all times. Therefore, FUV scans were only selected \textcolor{edits}{if DFMS was measuring} at similar times. The total neutral density at Rosetta was probed by the COmet Pressure Sensor (ROSINA/COPS) and has been corrected for the composition of the gas in line with \cite{Gasc2017PSS}. \newline
The neutral gas moves approximately radially away from the comet, which is along the line-of-sight when Alice is looking nadir. As a result, the whole column should originate from a similar region of the nucleus and have a constant composition throughout. \newline
The local neutral density measurements, $n(r_{Rosetta})$, at the cometocentric distance, $r_{Rosetta}$, of Rosetta can be converted to a radial column density from the nucleus surface to Rosetta as the neutral density has been observed to follow a $r^{-2}$-dependence \citep{Hassig2015, Bieler2015}:
\begin{equation}\label{eq: Haser column}
N_{Tot} =n(r_{Rosetta})r_{Rosetta}\bigg(\frac{r_{Rosetta}}{r_{\textcolor{edits}{\text{67P}}}}-1\bigg).
\end{equation} 
However, under this model, the total column density, $N_{Tot}$, is highly dependent on the radius of the comet, \textcolor{edits}{$r_\text{67P}$}, at the foot-print of the column, which varies significantly across the surface \citep{Jorda2016}. \textcolor{edits}{We have taken $r_\text{67P} = 1.7\text{~\si{\kilo\metre}}$ and assumed that the expansion velocity of the neutral gas is constant, which both introduce uncertainty to the neutral column density.} \textcolor{comments}{{The standard deviation of the cometoradius is approximately $0.26 \times r_{67P}$ \citep{Shap5Model}, which translates to a 30\% uncertainty in the column density, increasing to 35\% at low cometoradii ($\sim 10$km). The assumption of a constant expansion velocity results in an underestimate of the column density as the gas undergoes acceleration near the nucleus surface \citep{Heritier2017Vert, BykovZakharovOutflow2020}.}} When looking \textcolor{edits}{off-limb}, the neutral gas column probed by the FUV spectrograph \textcolor{edits}{may have very different properties to the gas measured locally by the pressure gauge, so the in situ neutral density was only used to derive the column density when looking close to nadir.} \newline
The column density of several of the major neutral species could be measured remotely using other instruments onboard Rosetta. The Visual InfraRed Thermal Imaging Spectrometer  \citep[\textcolor{edits}{VIRTIS,}][]{Coradini2007} observed emissions from the $\nu_3$ vibrational bands of \ce{H2O} and \ce{CO2} at 2.67~\si{\micro\metre} and 4.27~\si{\micro\metre} respectively \citep{Bockelee-Morvan2015}. The calculations used to derive the \ce{H2O} and \ce{CO2} column densities are the same as those outlined in \cite{Bockelee-Morvan2015}. Emission from the $\nu$(1-0) band of \ce{CO} was also observed by the IR spectrometer, but the emission in this band is much weaker than that of the $\nu_3$ bands of \ce{H2O} and \ce{CO2}, so the S/N was too low to be useable at the large heliocentric distances under focus here. VIRTIS comprised 2 channels, M and H, operating at similar wavelengths \textcolor{edits}{in the infrared} but VIRTIS-M stopped working in May 2015 \citep{Bockelee-Morvan2016}, so in the present study we used only VIRTIS-H measurements. \newline
The Microwave Instrument for Rosetta Orbiter \citep[MIRO; ][]{Gulkis2007} observed emissions from \ce{H2O} and \ce{CO} \textcolor{edits}{lines} in the sub-mm wavelengths. \textcolor{edits}{The (110-101) rotational lines of \ce{H2O} (\ce{H2^18O}) at 557~\si{\giga\hertz} (548~\si{\giga\hertz}) and the  \ce{CO} (5-4) rotational line at 576~\si{\giga\hertz} a were seen in emission spectra when observing the limb and in absorption spectra when viewing nadir \citep{Biver2015, Biver2019}. The \ce{CO} line was more difficult to observe due to its intrinsic low strength and the small abundance of \ce{CO}.} \newline
The IR and sub-mm spectrometers were aligned with the FUV spectrograph line-of-sight and \textcolor{edits}{their fields of view} were located close to Row 15 of the viewing slit. Thus, they probed approximately the same neutral column as the FUV spectrograph (near the centre of the slit). This is particularly useful when observing off-limb, as the composition may vary significantly along the column and the source of gas is far more dispersed. \textcolor{edits}{Column densities derived from MIRO data in nadir pointing are} less reliable when viewing the nucleus due to low contrast between the near surface warm gas emission and background radiation emitted from the surface. The IR spectrometer could be used when looking nadir if the surface is in shadow but \textcolor{edits}{it acquired little data in this configuration throughout the mission}. Each of these instruments has been used to constrain the column density when the data are available.
\subsection{Calculation of the emission frequency from dissociative excitation}\label{sec: emission freq}
The emission frequency, $\nu_l^X$, of a given neutral species, $l$, and emission line, $X$, is driven by only \textcolor{comments}{{two}} physical quantities (Eq. \ref{eq: model brightness}): the emission cross-section (see Section \ref{sec: Cross Sections}) and the suprathermal electron particle flux (see Section \ref{sec: electron flux}). 
\subsubsection{Dissociative excitation cross-sections}\label{sec: Cross Sections}

The emission cross-sections for each spectral line, $X$, and neutral species, $l$, are outlined in Table \ref{tab: Cross section outline}. The current set of laboratory measurements of the emission cross-sections due to electron impact are somewhat incomplete. Many of the cross-sections have datapoints at only one or two energies so \textcolor{edits}{the energy dependence of the cross-sections are not known.} As such, several assumptions about the \textcolor{edits}{energy dependence} of the cross-sections have been made and are summarised in Table \ref{tab: Cross section outline}. \newline
Several cross-sections were recently updated by \cite{Ajello2019}, but we do not use these in this study. The emission cross-sections of OI1356 from \ce{e + CO2} in \cite{Ajello2019} are an order of magnitude \textcolor{comments}{{smaller}} than those from previous literature \textcolor{comments}{{\citep{Wells1972, Wells1974}. The inferred OI1304/OI1356 line ratio of $\sim3.2$ is not consistent with emission spectra from the Rosetta mission, when electron impact on \ce{CO2} was prevalent \citep[Section \ref{sec: nadir cases} \& ][]{Feldman2018}. The $^5$S upper state of the OI1356 transition has a long radiative lifetime \citep[180~$\mu$s;][]{Wells1974} and may have been quenched through collisions. \cite{Wells1972} and \cite{Wells1974} measured the production and radiative lifetime of the $^5$S state independently, so the inferred emission cross section was not susceptible to collisions experienced by the intermediate state. The experiments in \cite{Ajello2019} were based on the emission spectra of \ce{CO2}, which are more strongly impacted by any quenching of the $^5$S state by the relatively dense neutral gas used ($\sim3\times 10^{11}$~molecules~cm$^{-3}$). This is 3-4 orders of magnitude denser than the neutral gas seen at 67P at the large heliocentric distances considered in this study.}} 
\begin{table*}[htbp]
    \centering
    \caption{Cross-sections for the dissociative excitation of cometary molecules by electron impact. \tablefoottext{1}{\cite{Mumma1972}}\tablefoottext{2}{\cite{McConkey2008}}.}
    \label{tab: Cross section outline}
    \begin{tabular}{||c|c|c|p{8cm}||}
    \hline
        Emission Line & Species & Threshold Energy [\si{\electronvolt}] & Reference and Assumptions \\
        \hline
        \multirow{4}{*}[-1.8em]{Ly-$\beta$ \& OI1027} & \ce{H2O} & 17.21 & \citet{Makarov2004}. Assume the same shape as Ly-$\alpha$ with ratio $7/55$ at 200~\si{\electronvolt}. Includes coincident OI line.\\
        & \ce{CO2} & $21\pm2$\tablefootmark{1} & \citet{Kanik1993} Assume the same shape as OI1304. Use ratio at 200~\si{\electronvolt}.\\
        & \ce{CO} & 23.17 & \citet{James1992}. \\
        & \ce{O2} & 20.6 & \citet{Wilhelmi2000}. Absolute value given at 200~\si{\electronvolt}. Assume same shape as OI1304. \\
        
        \hline
        \multirow{4}{*}[-1em]{OI1304} & \ce{H2O} & 15.2 & \citet{Makarov2004}.\\
        & \ce{CO2} & $21\pm2$ & \citet{Mumma1972}. Reduced by a factor of 0.59 due to updated measurements of \ce{H2} Ly-$\alpha$ \citep{McConkey2008}. \\
        & \ce{CO} & 20.6 & \citet{Ajello1971}. Some uncertainty at energies $>100$~\si{\electronvolt}.\\
        & \ce{O2} & 14.6\tablefootmark{2} & \citet{Kanik2003}. Scaled up by factor $2.93/2.90$ in line with the recommendation by \citet{McConkey2008}.  \\
        
        \hline
        
        \multirow{4}{*}[-2.5em]{OI1356} & \ce{H2O} & 15.2 & \citet{Makarov2004}. Assume same threshold and shape as OI1304. \\
        & \ce{CO2} & $20.6$\tablefootmark{2} & \citet{Feldman2015}. Excitation rate of OI1356 33 times larger for \ce{CO2} than \ce{H2O} \citep{Wells1972}, after revision of the lifetime of the \ce{^5S^$\circ$} state \citep{Wells1974}. Assume the same shape as OI1304. \\
        & \ce{CO} & 20.6 & \citet{Ajello1971}. Ratio given at 100~\si{\electronvolt} \citep{Wells1974, Wells1972}.  Assume the same shape as OI1304. \\
        & \ce{O2} & 14.6 & \citet{Kanik2003}. Scaled up by factor $6.47/6.40$ in line with recommendation by \citet{McConkey2008}.  \\
        
        \hline
        
        \multirow{2}{*}[-1em]{CI1657} & \ce{CO2} & $25\pm2$& \citet{Mumma1972}. \\
        &\ce{CO} & & Difficult to measure cross-section due to the strong overlapping CO4PG band. No contribution from dissociative excitation of CO considered for CI1657. \\
        \hline
        \multirow{2}{*}[-1em]{CO4PG} & \ce{CO2} & $25\pm2$ & Contribution of (0,2) bands included in CI1657 cross-section of \citet{Mumma1972}.\\
        & \ce{CO} & 8 & \citet{Beegle1999}. Absolute values for bands given at 100~\si{\electronvolt}. Shape of (0,1) band given. Scaled down by factor 0.925 due to remeasurement of calibrating NI line \citep{Ajello2019}.\\
        \hline
        \multirow{2}{*}{CII1335} & \ce{CO2} & 44 &\citet{Mumma1972}.\\
        & \ce{CO} & 33 & \citet{Ajello1971}.\\
        \hline
    \end{tabular}

\end{table*}
\subsubsection{Suprathermal electron flux}\label{sec: electron flux}
\begin{figure}[htbp]
\includegraphics[width = 0.5\textwidth]{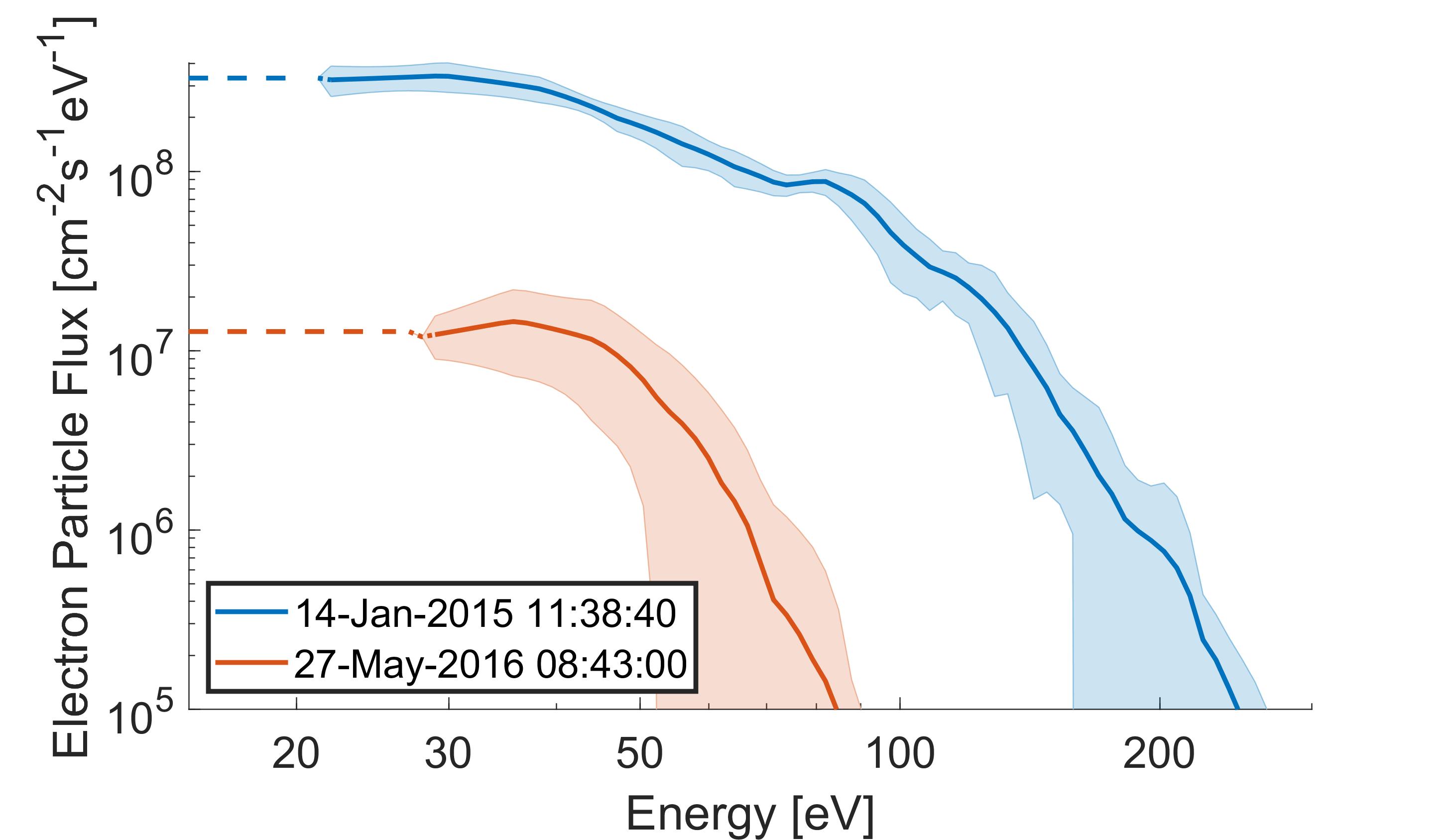}
\caption{Suprathermal electron particle flux during two nadir Alice scans during quiet periods. The solid line is the average particle flux during each of the scans, while the shaded region corresponds to the standard deviation of the electron flux in the same period. The extrapolation of the electron flux to energies below which RPC/IES could not measure, due to the spacecraft potential, is given by the dashed line.}
\label{fig: electron flux}
\end{figure}

The suprathermal electron flux during each scan of the FUV spectrograph has been derived from measurements by the Rosetta Plasma Consortium \citep[RPC;][]{Carr2007}. The count rate measurements from the Ion and Electron Sensor \citep[RPC/IES;][]{Burch2007} were converted to an electron particle flux using the method outlined in Appendix \ref{sec: L2 to L3}. \textcolor{edits}{Several of the RPC/IES anodes degraded throughout the mission, resulting in limited angular coverage of the electron flux \citep{Broiles2016a}. In the correction for the field of view, we assumed that the electron flux was isotropic.} \newline
\textcolor{edits}{We also assumed that the suprathermal electron flux was constant along the line of sight of the Alice FUV spectrograph, which is consistent with the findings of \citet{Chaufray2017}. We consider the electron depth \citep{Heritier2018a}
\begin{equation}\label{eq: elec depth}
\tau^{e-}=\sum\limits_l \sigma_l^{e-,inel}N_l(r),
\end{equation}
where $\sigma_l^{e-,inel}$ is the total inelastic collision cross-section for 30~\si{\electronvolt} electrons with the neutral species \ce{H2O} \citep[{$2.32\times 10^{-16}$}~\si{\square\centi\metre},][]{Itikawa2005}  and \ce{CO2} \citep[$1.6\times 10^{-16}$~\si{\square\centi\metre},][]{Itikawa2002}.
The electron depth is analogous to an optical depth, so for $\tau^{e-} <1$ we expect little degradation of suprathermal electrons along the line of sight. As expected at large heliocentric distances, the electron depth was small ($\tau^e<0.35$) for all cases in the present study so suprathermal electrons were unlikely to undergo collisions in the coma. }\newline
RPC/IES measured electrons that have energies between 4.32~\si{\electronvolt} and 17.67~\si{\kilo\electronvolt} at the detector. However, throughout the mission, the spacecraft was typically at a voltage of $-10$~\si{\volt}, as measured by the \textcolor{edits}{Rosetta Dual Langmuir Probes} \citep[\textcolor{edits}{RPC/LAP,}][]{Odelstad2015b}. The negative spacecraft potential ($V_{S/C}$) repelled electrons, allowing only those with higher energies to reach the sensor. The minimum energy, $E_{min}$, in \si{\electronvolt} that electrons observed by RPC/IES has is: $E_{min}\, \text{[\si{\electronvolt}]} = 4.32 - V_{S/C}\, \text{[\si{\volt}]}$. \newline
The electron flux was corrected for this using the Liouville's theorem, under the assumption that the phase space density is conserved within the potential of the spacecraft:
\begin{equation}\label{eq: Liouville}
\frac{J(E)}{E} = \frac{J_{IES}}{E_{IES}} \quad \text{where} \quad  E \, \text{[\si{\electronvolt}]}= E_{IES} \, \text{[\si{\electronvolt}]}- qV_{S/C}\, \text{[\si{\volt}].}
\end{equation}
Within the duration of each FUV scan, there were several measurements of the electron flux by RPC/IES, each of which was individually corrected for the spacecraft potential at that time. The time-average and the standard deviation of the electron flux were calculated at each energy as shown in Figure \ref{fig: electron flux}. \newline
Although RPC/IES could not measure the electron population below the detection threshold, there may still have been a large electron flux at low energies. As such, we extended the mean particle flux to low energies assuming a constant particle flux at low energies. We also considered extrapolating logarithmically but the method of extrapolation had little impact on the resulting emission frequency as the cross-sections decrease significantly near the threshold energies (given in Table \ref{tab: Cross section outline}).
\textcolor{edits}{The emission cross-sections decrease sharply near the threshold energies of each line (see Table~\ref{tab: Cross section outline}) and electrons below the threshold are unable to generate FUV emissisons. Despite large particle fluxes the low energy ($<20$~\si{\electronvolt}) electrons contributed little to the model brightness.}

\subsection{Other sources of emission} \label{sec: Other Sources}
Alongside dissociative excitation by electron impact, there are several other sources of emission that are observed at comets. We have already referred to the contribution to the Lyman series by the IPM (see Section \ref{sec: Alice Obs}), but this should not have been visible when viewing the surface of the nucleus from Rosetta. \newline
We considered two main emission sources outside of electron impact: prompt-photodissociation of \ce{H2O} to produce Lyman-$\beta$ and fluorescence of CO to emit in the Fourth Positive Group. \newline
These emission features are driven by the solar flux incident on the column of gas along the line of sight. The brightness of the emissions from these sources is given by
\begin{equation}\label{eq: solar emission}
B^{X, h\nu}_l = 10^{-6} N_l \int\limits_{\lambda_{min}}^{\lambda_{Th}}\! \sigma_{l}^{X,h\nu}(\lambda)I(\lambda)\,\mathrm{d}\lambda,
\end{equation}
where $l$ is \ce{H2O} for $X=\text{Ly-$\beta$}$ and \ce{CO} for $X=\text{CO4PG}$ (contributing to the CI1657 emissions).\newline
The photon flux, $I(\lambda)$, at comet 67P was driven by TIMED-SEE \citep{Woods2005} measurements at 1~\si{\astronomicalunit} \textcolor{edits}{taken at the same Carrington longitude as comet 67P for each time interval}.
The cross-section for resonance fluorescence of \ce{CO} was based on the model of \cite{Lupu2007}. For the prompt-photodissociation of \ce{H2O}, we use the emission cross-section from  \cite{Hans2015}.  \newline
The modelled brightness of these photon-driven emissions are upper bounds as we assumed that the entire column of neutral gas along the line of sight was illuminated. In reality, the neutral column may have been partially shadowed near the nucleus, where the neutral number density was highest, and the emissions from fluorescence of CO and prompt-photodissociation of \ce{H2O} will be overestimated.
\section{Nadir analysis during quiet periods} \label{sec: Nadir}
\subsection{Selected cases}\label{sec: nadir cases}
\begin{figure*}[htbp]

    \centering
    \begin{tikzpicture}[node distance= 1cm, baseline=(current  bounding  box.center)]
    \node (emfreq) at (0,0) {\includegraphics[width = \textwidth]{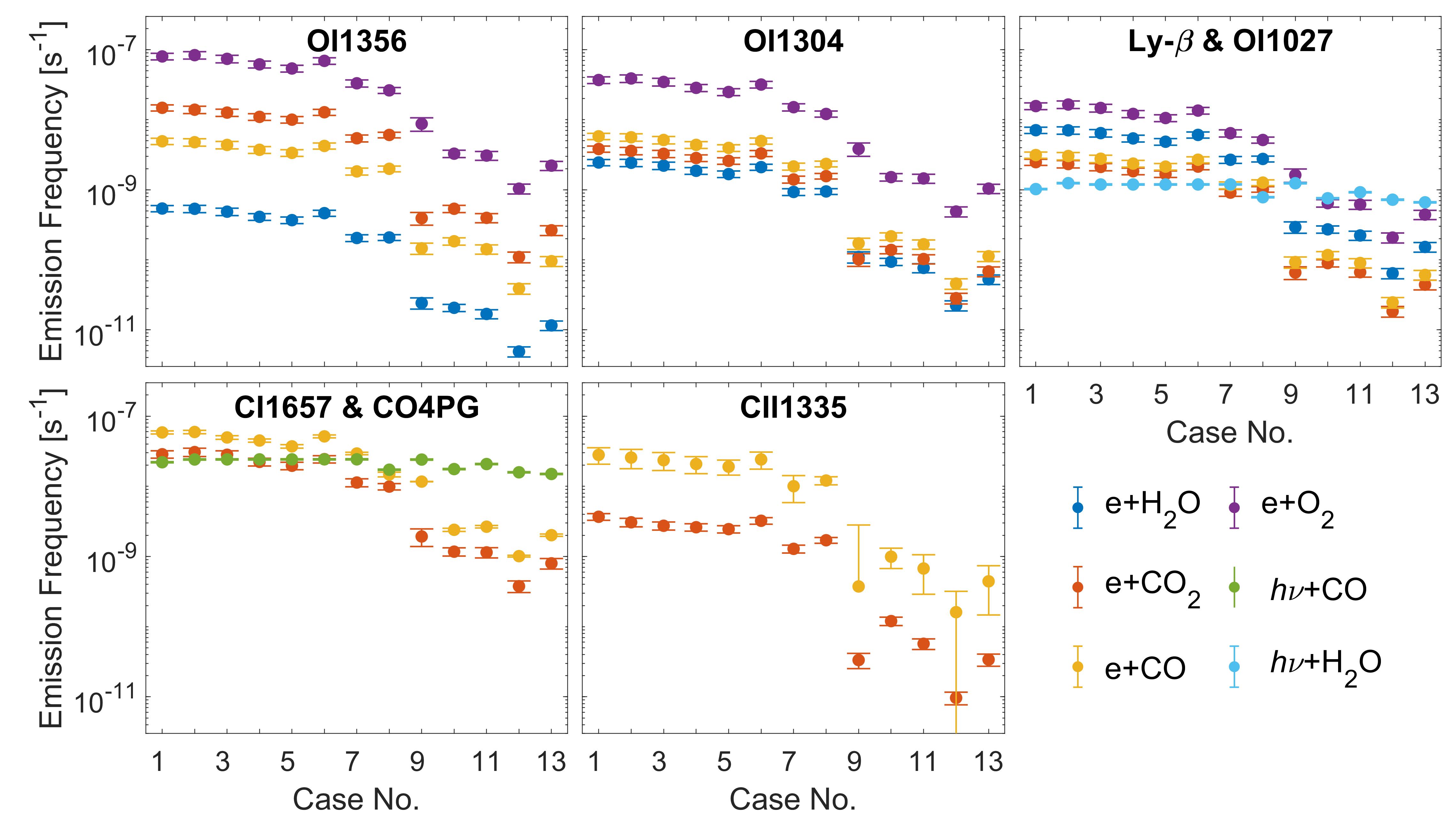}};
    \node [above left = 5.9 and 6.5 of emfreq.base, font =\large] (A) {(a)};
    \node [above left = 5.9 and 1 of emfreq.base, font =\large] (B) {(b)};
    \node [above right = 5.9 and 4 of emfreq.base, font =\large] (C) {(c)};
    \node [above left = 1.4  and 6.5 of emfreq.base, font =\large] (D) {(d)};
    \node [above left = 1.4 and 1 of emfreq.base, font =\large] (E) {(e)};
    \end{tikzpicture}

    \caption{Emission frequency, $v_l^X$, of each neutral species and emission line: (a) OI1356, (b) OI3014, (c) Ly-$\beta$ and OI1027, (d) CI1657 and CO4PG, and (e) CII1335. Emissions due to electron impact (\ce{e + X}, Section \ref{sec: emission freq}) on \ce{CO2} (red), \ce{H2O} (dark blue), CO (orange), and \ce{O2} (purple) and other processes (\ce{h{\nu} + X}, Section \ref{sec: Other Sources}) have been included. The uncertainty in the electron impact emission frequency is derived from the variability in the electron flux during each scan of the FUV spectrograph (see Section \ref{sec: electron flux}). \ce{h{\nu} + CO} refers to emissions from the fluorescence of CO (green), while \ce{h{\nu} + H2O} refers to the prompt-photodissociation of water (light blue) to produce Lyman-$\beta$ (see Section \ref{sec: Other Sources}). }
    \label{fig: emission freq}

\end{figure*}
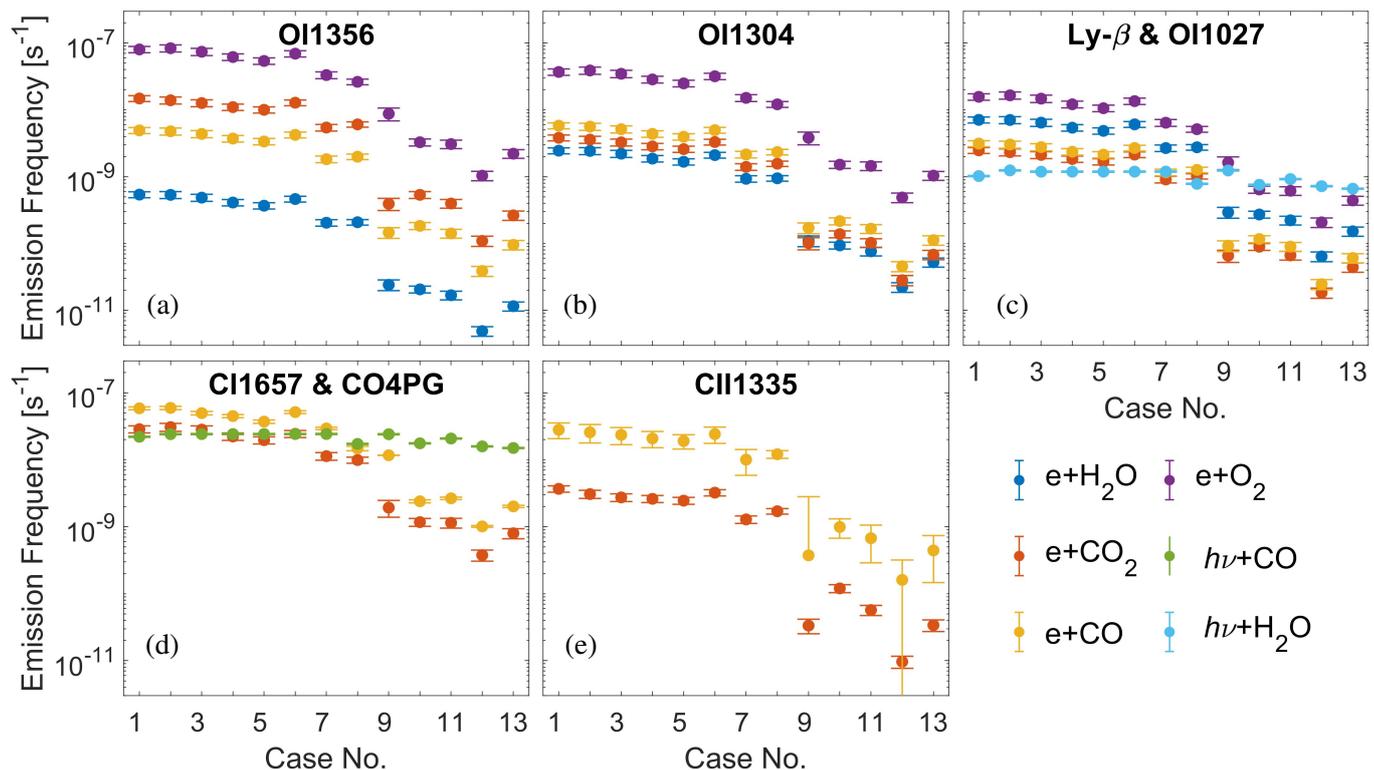

\begin{figure}[htbp]
    \includegraphics[width = 0.5\textwidth]{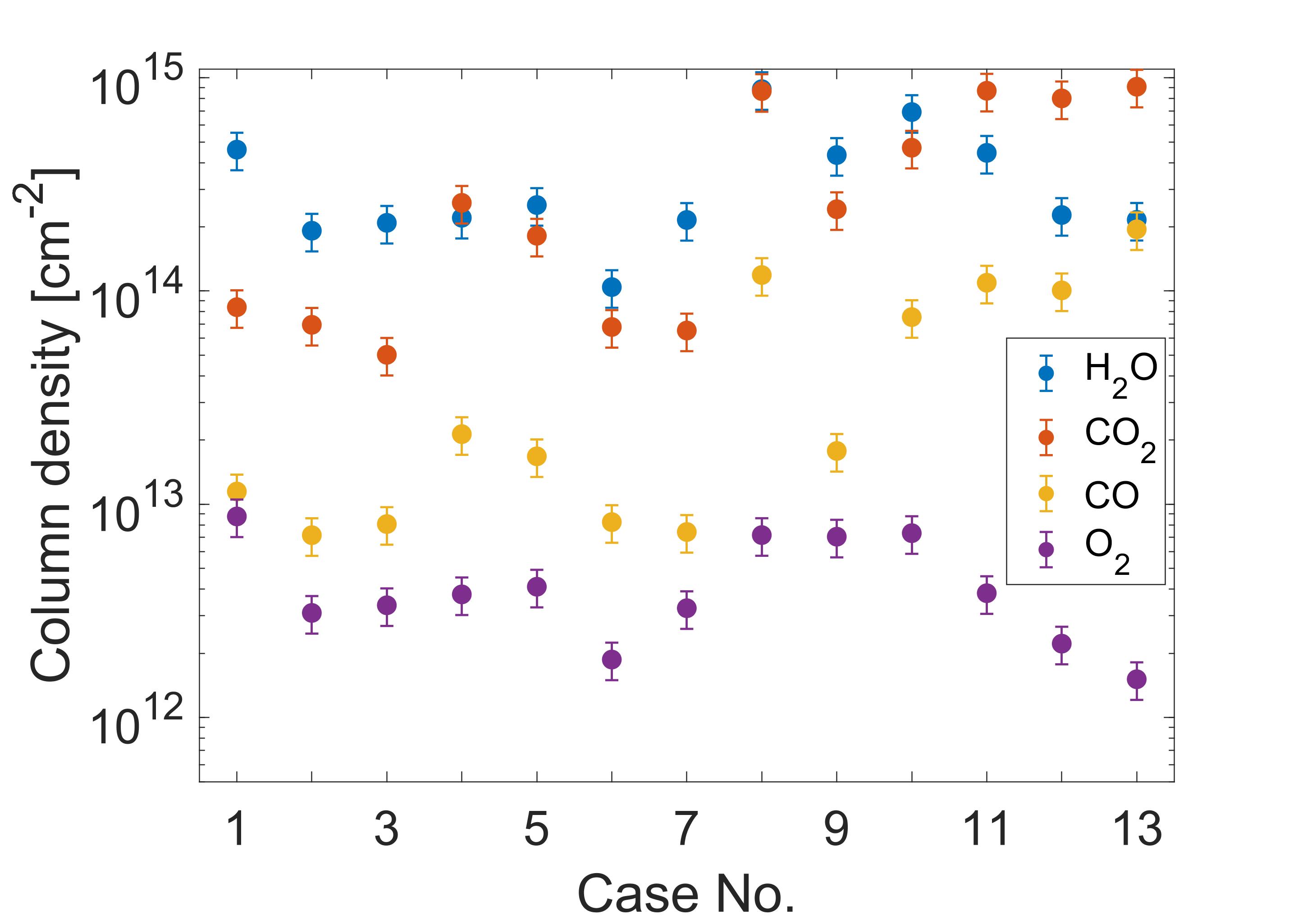}
    \caption{Column density of the four major neutral species in the coma during each of the spectrograph scans. \textcolor{edits}{The volume mixing ratios are derived from ROSINA/DFMS measurements (see Section \ref{sec: Neutral Column}).}}
    \label{fig: column density}
\end{figure}

\begin{table*}[htbp]
    \centering
    \caption{Cases selected in the southern hemisphere, with nadir viewing at large heliocentric distances. The horizontal lines separate nadir cases with high (Cases 1-8) and low (Cases 9-13) electron fluxes.}
    \label{tab: Nadir cases}
    {\tabulinesep = 1.2mm 
    \begin{tabular}{||>{\centering\arraybackslash}m{0.6cm}|>{\centering\arraybackslash}m{1.5cm}|>{\centering\arraybackslash}m{1.3cm}|>{\centering\arraybackslash}m{1.5cm}|>{\centering\arraybackslash}m{2cm}|>{\centering\arraybackslash}m{1.9cm}|>{\centering\arraybackslash}m{1.5cm}|>{\centering\arraybackslash}m{1.5cm}|>{\centering\arraybackslash}m{2.2cm}||}
    
    \hline
         \thead{Case} & \thead{Date} & \thead{Start \\ Time \\ {[UTC]}} & \thead{Integration \\ Time \\ {[\si{\second}]}} & \thead{Cometocentric \\ Distance \\ {[\si{\kilo\metre}]}} & \thead{Heliocentric \\ Distance \\ {[\si{\astronomicalunit}]}}
         & \thead{Spacecraft \\ Latitude \\ {[$\circ$]}} & \thead{Spacecraft \\ Longitude \\ {[$\circ$]}} & \thead{Total Column \\ Density \\ {[$10^{15}$~\si{\per\square\centi\metre}]}} \\
         \hline
         1 & 14/01/2015 & 11:38:40 & 3629 & 28.4 & 2.55 & -23.4 & -82.9 & 0.58 \\
         2 & 29/01/2015 & 18:40:49 & 6048 & 27.8 & 2.44 & -63.8 & 5.4 & 0.27 \\
         3 & 30/01/2015 & 04:25:44 & 2419 & 27.8 & 2.43 & -64.6 & 56.6 & 0.22 \\
         4 & 30/01/2015 & 06:34:42 & 2419 & 27.8 & 2.43 & -64.2 & -11.4 & 0.51 \\
         5 & 30/01/2015 & 07:17:41 & 1763 & 27.8 & 2.43 & -64.0 & -34.0 & 0.48 \\
         6 & 30/01/2015 & 11:32:18 & 2419 & 27.9 & 2.43 & -62.5 & -167.4 & 0.18 \\
         7 & 30/01/2015 & 15:25:50 & 2419 & 27.9 & 2.43 & -60.5 & 71.35 & 0.28 \\
         8 & 26/04/2016 & 06:06:12 & 4452 & 21.2 & 2.9 & -33.1 & -107 & 1.88 \\
         \hline
         9 & 29/01/2015 & 07:51:27 & 3629 & 27.8 & 2.44 & -58.4 & -17.1 & 0.75 \\
         10 & 21/04/2016 & 23:01:00 & 3398 & 30.9 & 2.85 & -24.6 & -57.7 & 1.24 \\
         11 & 21/03/2016 & 00:05:00 & 2603 & 12.2 & 2.62 & -24.2 & -36.8 & 1.43 \\
         12 & 14/05/2016 & 14:39:34 & 4613 & 9.8 & 3.0 & -53.1 & -111.7 & 1.13 \\
         13 & 27/05/2016 & 08:43:00 & 5592 & 7.0 & 3.09 & -56.2 & -42.3 & 1.32 \\
         
         \hline
         
    \end{tabular}
    }
    
\end{table*}
In order to determine whether the FUV emissions over the southern hemisphere are driven primarily by electron impact, cases with a nadir viewing geometry over the shadowed nucleus are considered. By observing the shadowed nucleus, there is no contribution from the IPM to the observed Lyman-$\beta$ brightness and any contamination from solar photons reflected off the nucleus is minimised. This geometry also provides the best constraint on the neutral composition of the coma from the mass spectrometer (see Section \ref{sec: Neutral Column}). \newline
We have considered 13 cases in the southern hemisphere and at large heliocentric distances. The properties of each of these scans are outlined in Table \ref{tab: Nadir cases}. The cases have been split into two sets, which have been approached separately: nadir cases with a high suprathermal electron flux, as illustrated by the blue spectrum in Fig.~\ref{fig: electron flux} (Cases 1-8; Section \ref{sec: nadir 1-8}); and nadir cases with a low suprathermal electron flux, as illustrated by the red spectrum in Fig.~\ref{fig: electron flux} (Cases 9-13; Section \ref{sec: nadir 9-13}). The emission frequency for all the selected lines and the column density of each neutral species are shown in Figs.~\ref{fig: emission freq} and \ref{fig: column density}, respectively, for all cases in order to interpret the modelled brightnesses presented in Fig.~\ref{fig: Nadir Quiet}, along with the FUV spectrograph observations. 
The average electron particle fluxes in two energy brackets are given in Table \ref{tab: electron flux}. The distinction between the high and low suprathermal electron flux cases (see Fig.~\ref{fig: electron flux}) can be seen in both of the energy ranges, but to a greater extent from $60-120$~\si{\electronvolt}. \newline
\begin{table}[htbp]
\centering
\caption{Average electron flux in two energy ranges for each of the nadir cases \textcolor{edits}{outlined in Table \ref{tab: Nadir cases}}. The distinction between cases with high and low suprathermal electron fluxes is clearer in the 60-120~\si{\electronvolt} range.}
\label{tab: electron flux}
    \begin{tabular}{||>{\centering\arraybackslash}m{0.6cm}|>{\centering\arraybackslash}m{3cm}|>{\centering\arraybackslash}m{3cm}||}
    \hline
         Case & \makecell{Ave. Electron Flux \\ for $20 - 60$~\si{\electronvolt} \\{[$ 10^7$~\si{\per\square\centi\metre\per\second\per\electronvolt}]}}
         & \makecell{Ave. Electron Flux \\ for $60 - 120$~\si{\electronvolt}\\ {[$10^7$ \si{\per\square\centi\metre\per\second\per\electronvolt}]}} \\
         \hline
         1 & $23.2 \pm 1.2$ & $6.33 \pm 0.85$ \\
         2 & $30.4 \pm 1.9$ & $5.68\pm 1.34$ \\
         3 & $27.7 \pm 3.6$ & $5.10\pm 2.62$ \\
         4 & $20.5 \pm 1.3$ & $4.82 \pm 0.95$ \\
         5 & $16.9\pm 0.9$ & $4.36\pm0.65$ \\
         6 & $22.2 \pm 0.8 $ & $5.88\pm 0.56$ \\
         7 & $10.5 \pm 1.4$ & $2.47 \pm 1.01$ \\
         8 & $\phantom{0}6.42\pm 0.51$ & $2.60\pm 0.37$ \\
         \hline
         9 & $3.51 \pm 0.84$ & $0.05\pm 0.61$ \\
         10 & $1.18 \pm 0.05$ & $0.22\pm 0.04$ \\
         11 & $1.40 \pm 0.12$ & $0.06 \pm 0.09$ \\
         12 & $0.51 \pm 0.09$ & $0.01 \pm 0.07$ \\
         13 & $1.03 \pm 0.11$ & $0.03 \pm 0.08$ \\
         \hline
    \end{tabular}
    
\end{table}

\subsection{Nadir cases 1-8}\label{sec: nadir 1-8}
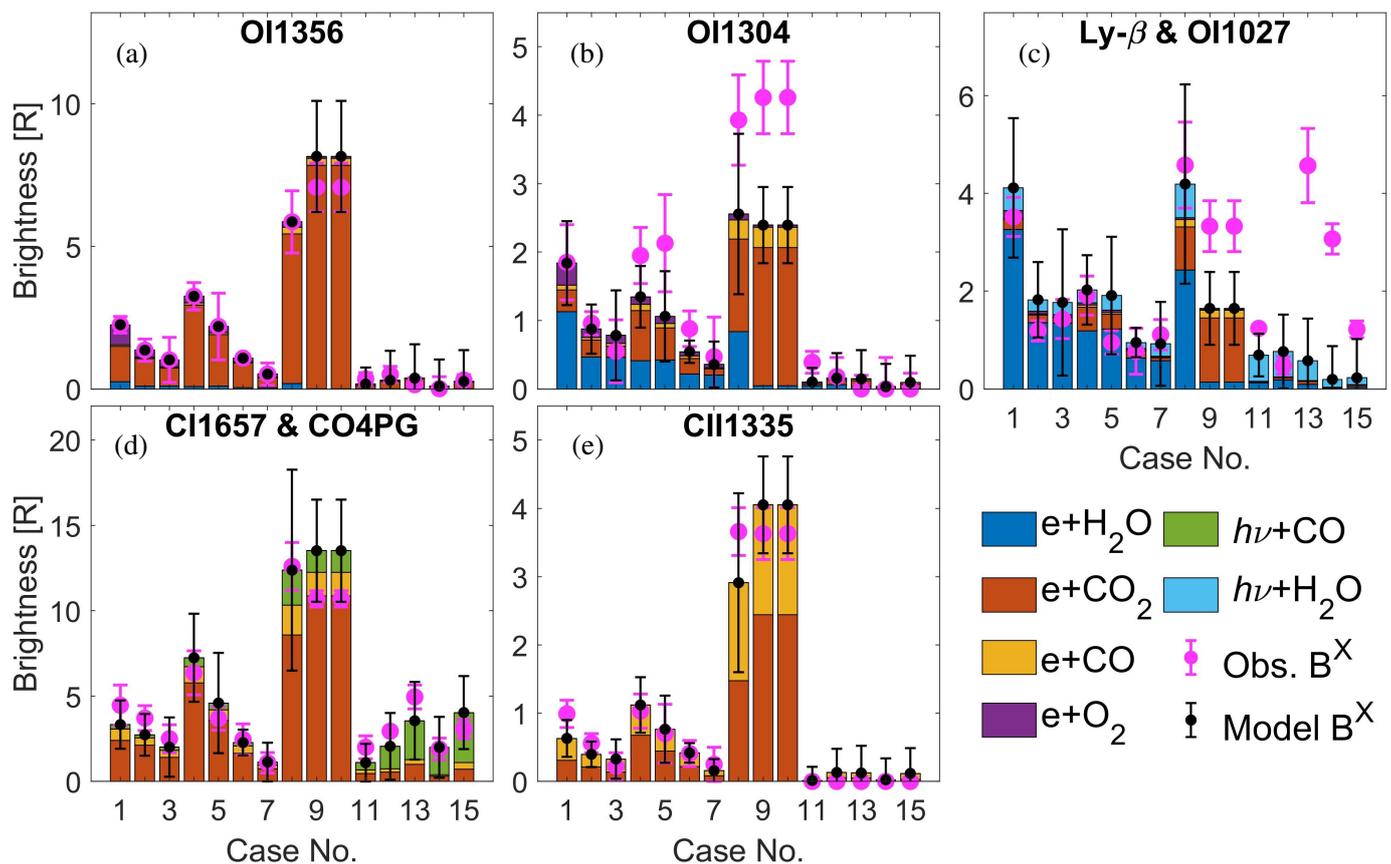
\begin{figure*}[htbp]
    \centering
    \begin{tikzpicture}[node distance= 1cm, baseline=(current  bounding  box.center)]
    \node (nadir) at (0,0) {\includegraphics[width = \textwidth]{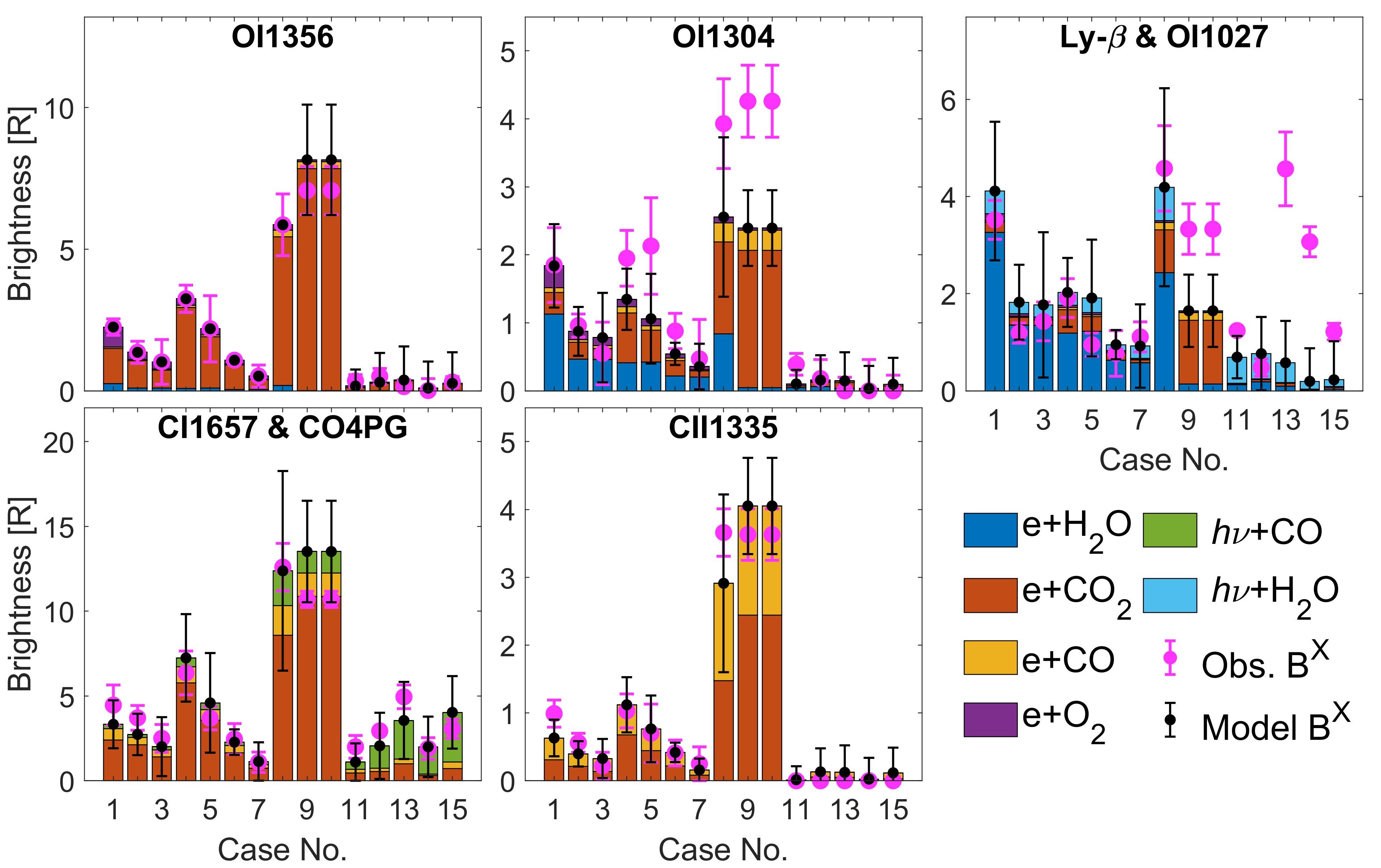}};
    \node [above left = 10.5 and 7.2 of nadir.base, font =\large] (A) {(a)};
    \node [above left = 10.5 and 1.2 of nadir.base, font =\large] (B) {(b)};
    \node [above right = 10.5 and 4 of nadir.base, font =\large] (C) {(c)};
    \node [above left = 5.3 and 7.2 of nadir.base, font =\large] (D) {(d)};
    \node [above left = 5.3 and 1.2 of nadir.base, font =\large] (E) {(e)};
    \end{tikzpicture}
    
    \caption{Comparison of the total modelled (black) and observed brightness (magenta), $B^X$, of each emission line. The stacked bars show the contribution from each neutral species and emission process. The same colour code as Fig.~\ref{fig: emission freq} is used. The error on the observed brightness is derived from the integration of the FUV spectra. The error on the modelled brightness is from the temporal variation of the electron flux and column density, as well as a 20\% uncertainty in the neutral composition. \textcolor{edits}{In cases 1-8, where the observed OI1356 brightness is used to fit the total neutral column density (see Section \ref{sec: nadir 1-8}), the error on the modelled OI1356 brightness is included in the error of the other modelled emission brightnesses.}}
    \label{fig: Nadir Quiet}

\end{figure*}

As mentioned in Section \ref{sec: Neutral Column}, we can derive the total neutral density in nadir viewing from in situ measurements. However, the strong dependence of the column density on the cometoradius means this method is quite uncertain. Alternatively, we can constrain the total column density by setting the modelled brightness of the OI1356 line such that it equals the observed brightness of this line. OI1356 emissions are associated with a forbidden transition (\ce{^5S - ^3P}), so there are few sources of this emission line apart from electron impact. The contributions of both resonance scattering and fluorescence to this line are negligible. \newline
There is a close agreement between the modelled and observed brightnesses for all five FUV emission lines selected for cases 1-8 (see Fig.~\ref{fig: Nadir Quiet}). This suggests that our model represents the sources of FUV emissions in the coma well. \newline
The emission of OI1356 is dominated by \ce{e + CO2} (red, Fig.~\ref{fig: Nadir Quiet}a) across the high suprathermal electron flux cases. This is a result of the high emission frequency of \ce{e + CO2} ($1.08\times 10^{-8}$~\si{\per\second}) compared to \ce{e + H2O} ($4.0\times10^{-10}$~\si{\per\second}, see Fig.~\ref{fig: emission freq}) in this line.  The small emission frequency from \ce{e + H2O} means the brightness of the OI1356 emissions is not sensitive to the column density of water. \ce{e + O2} has an OI1356 emission frequency of $5.98\times10^{-8}$~\si{\per\second} in cases 1-8, larger than that from \ce{e + CO2}, but the column density of \ce{O2} is insufficient for this process to contribute significantly to the total brightness (purple, Fig.~\ref{fig: column density}). In case 1, \ce{e + O2} drives 0.69~R of OI1356 emission (purple, Fig.~\ref{fig: Nadir Quiet}a), which is considerably more than the $0.1-0.2$~R of emission from this source in cases 2-8. This results from the larger column density of \ce{O2} ($8.9\times10^{12}$~\si{\per\square\centi\metre}, Fig.~\ref{fig: column density}) in case 1 than in cases 2-7 ($3.1\pm0.9 \times 10^{12}$~\si{\per\square\centi\metre}, Fig.~\ref{fig: column density}), due to the more equatorial latitude compared to the southerly cases 2-7 (see Table \ref{tab: Nadir cases}). Case 8 occurred post-perihelion, when the outgassing rate of \ce{CO2} increased, resulting in a low volume mixing ratio of \ce{O2} (purple, Fig.~\ref{fig: column density}), despite having a similar latitude to case 1 (see Table \ref{tab: Nadir cases}).  \newline
The OI1304 emissions are mostly driven by electron impact on \ce{CO2} and \ce{H2O} (see Fig.~\ref{fig: Nadir Quiet}b). The emission frequency of OI1304 from \ce{e + CO2} is only 1.5 times more efficient than \ce{e + H2O}, in contrast to the factor of 20 in the emission frequencies of OI1356. Electron impact on \ce{CO} and \ce{O2} are 1.5 and 10 times more efficient at emitting OI1304 than \ce{e + CO2} (see Fig.~\ref{fig: emission freq}b), respectively, but the small volume mixing ratios of these molecules throughout cases 1-8 ($\ce{O2}/\ce{CO2}=0.04$; $\ce{CO}/\ce{CO2}=0.11$; Fig.~\ref{fig: column density}) limits their contribution to the total OI1304 brightness. However, these processes can be a significant source of OI1304, when the volume mixing ratio of each species increases (see. Fig.~\ref{fig: column density}) as seen in case 1 for \ce{e + O2} and case 8 for \ce{e + CO} (see Fig.~\ref{fig: Nadir Quiet}b). \newline
The emission feature near 1026~\si{\angstrom} is dominated by electron impact on water throughout cases 1-8 (dark blue, Fig.~\ref{fig: Nadir Quiet}c), generating emissions of Ly-$\beta$. As seen with OI1356 and OI1304 emissions, the \textcolor{comments}{{largest emission frequency at this wavelength is from}} \ce{e + O2} at $1.17\times10^{-8}$~\si{\per\second} (purple, Fig.~\ref{fig: emission freq}c), which produces OI1027, but this is only twice that of \ce{e + H2O}, so the small column density of \ce{O2} with respect to that of water means it contributes negligibly to the modelled brightness (purple, Fig.~\ref{fig: Nadir Quiet}). Throughout these cases, we see a sizeable contribution from \ce{e + CO2} ($<0.9$~R) to the OI1027 brightness that is not seen in the northern hemisphere \citep{Galand2020}. This is especially prominent in case 8, with a post-perihelion enhanced \ce{CO2} column density of $8.66\times 10^{14}$~\si{\per\square\centi\metre} (see Fig.~\ref{fig: column density}). In cases 1-8, prompt-photodissociation of \ce{H2O} (light blue, Fig.~\ref{fig: emission freq}c) has an emission frequency 5 times smaller than from  \ce{e + H2O} (dark blue, Fig.~\ref{fig: emission freq}c), so photodissociation is a minor source of emissions when the suprathermal electron flux is large (Fig.~\ref{fig: Nadir Quiet}c). We show that \ce{e + H2O} is the major source of the emissions near 1026~\si{\angstrom} in the southern hemisphere, but the contribution from \ce{e + CO2} and prompt-photodissociation of \ce{H2O} to this line are not negligible. \newline
Throughout cases 1-8, the emissions near 1657~\si{\angstrom} are dominated by CI1657 emission from electron impact on \ce{CO2} (red, Fig.~\ref{fig: Nadir Quiet}d). There is also a small contribution from \ce{e + CO} (orange, Fig.~\ref{fig: Nadir Quiet}d) to the overlapping bands of the Fourth Positive Group, which has an emission frequency ($4.58\times10^{-8}$~\si{\per\second}) twice that of \ce{e + CO2} (orange and red, respectively, Fig \ref{fig: emission freq}d). As the suprathermal electron flux is high for these cases, fluorescence of CO (green, Fig.~\ref{fig: emission freq}d) has a lower or similar emission frequency to \ce{e + CO2}. 
However, the low column density of \ce{CO} compared to that of \ce{CO2} (see Fig.~\ref{fig: column density}) means the emissions from both electron impact on and fluorescence of \ce{CO} are only minor contributions to the total brightness near 1657~\si{\angstrom} (see Fig.~\ref{fig: Nadir Quiet}d), except in case 8, when the column density of \ce{CO} ($1.19\times10^{14}$~\si{\per\square\centi\metre}) is larger than seen in cases 1-7 (see Fig.~\ref{fig: column density}). \newline
The CII1335 emissions have significant contributions from electron impact on both \ce{CO} and \ce{CO2}. The emission frequency of \ce{e + CO} ($2.04\times 10^{-8}$~\si{\per\second}) is 8 times larger from \ce{e + CO2} for this line, which is balanced by a ratio of 0.11 of \ce{CO} to \ce{CO2} in column density. The two competing effects result in the two species contributing approximately equally to the brightness of the CII1335 line (see Fig.~\ref{fig: Nadir Quiet}e). This emission line is primarily driven by $60-120$~\si{\electronvolt} electrons due to the high threshold energies of \ce{e + CO} (33~\si{\electronvolt}) and \ce{e + CO2} (44eV, Table \ref{tab: Cross section outline}) for this process. The electrons in this energy range vary little across cases 1-8 (see Table \ref{tab: electron flux}), which is reflected in the roughly constant emission frequency (see Fig.~\ref{fig: emission freq}e). Therefore, the variations in the CII1335 brightness in cases 1-8 are driven by changes in the column densities of \ce{CO} and \ce{CO2} (see Fig.~\ref{fig: column density}). The large emission frequency of \ce{e + CO} relative to \ce{e + CO2} (see Fig.~\ref{fig: emission freq}e) in this line means the observed brightness is highly sensitive to the column density of \ce{CO}. The volume mixing ratio of CO, derived from measurements by the mass spectrometer, includes a significant correction for fragmentation of \ce{CO2} within the instrument, which leads to a contribution to the CO signal \citep{Dhooghe2014}. The close agreement between the modelled and observed brightness of this line across cases 1-8 (Fig.~\ref{fig: Nadir Quiet}e) suggests that the corrected \textcolor{edits}{$\ce{CO}/\ce{CO2}$} volume mixing ratio derived from the mass spectrometer measurements is accurate.
\subsection{Nadir cases \textcolor{comments}{{9-13}}}\label{sec: nadir 9-13}
In cases \textcolor{comments}{{9-13}}, the suprathermal electron flux was very low (see Table \ref{tab: electron flux}) in both energy ranges. The electron flux was on average 11.8 times greater for cases 1-8 than for cases 9-13 \textcolor{edits}{between 20 and 60~\si{\electronvolt}}, whereas in the range $60-120$~\si{\electronvolt} the average flux ratio was \textcolor{edits}{$\sim\!{60}$}. The FUV emission lines considered, except CII1335, are driven primarily by electrons from $20-60$~\si{\electronvolt} so the emission frequencies are a factor of $\sim\!20$ smaller in cases 9-13 than in cases 1-8 (e.g. 26.6 for Ly-$\beta$ from \ce{e + H2O}; dark blue, Fig.~\ref{fig: emission freq}c). The threshold energies for emission of CII1335 from electron impact on \ce{CO} (33~\si{\electronvolt}, see Table \ref{tab: Cross section outline}) and \ce{CO2} (44~\si{\electronvolt}) are much higher than for the other wavelengths considered, and hence the emissions are more dependent on the electron flux between 60 and 120~\si{\electronvolt}. This results in a ratio of \textcolor{edits}{$\sim\!{50}$} in the emission frequency of the process \ce{e + CO2} (red, Fig.~\ref{fig: emission freq}e) between the high and low flux cases. \newline
Throughout the low flux cases, we observe no substantial emissions of OI1356 due to the low electron impact emission frequencies (see Fig.~\ref{fig: Nadir Quiet}a). As a result, we use the radial column density, derived from the in situ pressure gauge measurements (see Section~\ref{sec: Neutral Column}), to calculate the modelled emission brightnesses. \textcolor{edits}{The radial column density is calculated assuming a constant neutral gas velocity, which may result in an underestimate of the column density.} \newline
We also observe negligible emissions of OI1304 throughout this time (see Fig.~\ref{fig: Nadir Quiet}b). The column densities in cases 9-13 are similar to cases 1-8 (see Fig.~\ref{fig: column density}), so the lack of oxygen line emissions is a result of the lower electron flux (Table \ref{tab: electron flux}). In addition, there are negligible emissions of CII1335 (see Fig.~\ref{fig: Nadir Quiet}e), as there are few $60-120$~\si{\electronvolt} electrons in the coma (Table \ref{tab: electron flux}). The lack of observed OI1356, OI1304, and CII1335 emissions (see Figs.~\ref{fig: Nadir Quiet}a, \ref{fig: Nadir Quiet}b and ~\ref{fig: Nadir Quiet}e) in these lines is replicated in the modelled brightnesses for the cases associated with a low suprathermal electron flux. This demonstrates that the only source of these emission lines in nadir viewing is dissociative excitation by electron impact. \newline
As shown in Fig. ~\ref{fig: Nadir Quiet}d, there are significant emissions near 1657~\si{\angstrom} for these cases, despite the small populations of energetic electrons in the coma (Table \ref{tab: electron flux}) and the low emission frequency of the processes \ce{e + CO} ($3.96\times10^{-9}$~\si{\per\second}) and \ce{e + CO2} ($1.07\times10^{-9}$~\si{\per\second}, see Fig.~\ref{fig: emission freq}d). The emission frequency of fluorescence of CO is $1.86\times10^{-8}$~\si{\per\second} through cases 9-13, and hence is the dominant source of emissions at this wavelength. The dominance of \ce{CO} fluorescence, compared to case \textcolor{comments}{{8}}, results only from the decrease in the electron impact emission frequency, as both the volume mixing ratio $\ce{CO}/\ce{CO2} = 0.14\pm0.05$ and the emission frequency from CO fluorescence (green, Fig.~\ref{fig: Nadir Quiet}d) are similar to those in case 8. \textcolor{comments}{{In cases 1-7, the total column density and the CO column density are on average a factor 3 times smaller than in cases 9-13, which result in a small absolute contribution from CO fluorescence (0.27~R).}} \textcolor{comments}{{There is some uncertainty in the source of the CO4PG emissions in the low flux cases, as the Cameron bands, which are seen at long wavelengths, indicate that there may be emissions from CO in the 4PG driven by a large flux of low energy ($\sim$10~eV) electrons. However, we have not seen evidence of this in the measured electron flux as the spacecraft potential ($-23$ to $-17$~eV, as measured by RPC/LAP) prevented measurements at such low electron energies. Other emission lines are unaffected by these low energy electrons as these lines have a much higher threshold energy ($>$15~eV) compared to 4PG emissions from CO (8~eV, see Table \ref{tab: Cross section outline}).}} Despite this, the modelled CI1657 and CO4PG emissions well represent the observed emissions in these cases, given the uncertainty in the column density, and dissociative excitation by electron impact is not the major source of emissions when the suprathermal electron flux is small. \newline
In cases 9-13, the emission frequency of Ly-$\beta$ and OI1027 from electron impact on all species drops below the emission frequency of prompt-photodissociation of \ce{H2O} ($8.6\times10^{-10}$~\si{\per\second}, Fig.~\ref{fig: emission freq}c). In these cases, prompt-photodissociation of \ce{H2O} is the dominant source of emissions as the process \ce{e + H2O} has an emission frequency a factor four smaller ($1.97\times10^{-10}$~\si{\per\second}, Fig.~\ref{fig: emission freq}c). Photodissociation of \ce{H2O} is comparable in efficiency to electron impact on \ce{O2} near 1026~\si{\angstrom} but throughout these cases the number density ratio of $\ce{O2}/\ce{H2O}$ is less than $0.02$, so the contribution from electron impacts on \ce{O2} is small. \newline
The observed and modelled brightnesses of Ly-$\beta$ and OI1027 agree closely in cases 9 and 10 but for cases 11-13 the modelled brightness is smaller than that observed (see Fig.~\ref{fig: Nadir Quiet}c). \textcolor{comments}{{The unexplained intense Lyman-$\beta$ emissions are unlikely to be from electron impact emissions which would produce strong signals in the other atomic lines, such as OI1304 from \ce{e + H2O} at a factor $\sim 2.9$ smaller than the Ly-$\beta$ brightness. \newline
The Lyman-$\alpha$/Lyman-$\beta$ ratio can be used as an indicator of key emission processes, with a ratio of 8 expected from pure \ce{e + H2O} \citep{Makarov2004}. On 29 Nov 2014, a spectrum of pure \ce{e + H2O} \citep{Galand2020} was observed with a ratio Lyman-$\alpha$/Lyman-$\beta = 4.4$. The disparity between the observed ratio and the theoretical ratio may be a result of the difficult calibration of Alice around Lyman-$\alpha$, variation in the shape of the Lyman series cross-sections or differences in the cascade emissions. Case 8, also in early 2016, has a ratio of 2.56 with a large electron flux, but the 1027{\AA} line brightness has a large contribution of OI emissions from \ce{e + CO2}. The modelled Ly-$\beta$ emissions from \ce{e + H2O} in case 8 are a factor 4.4 smaller than the Ly-$\alpha$ brightness, consistent with the line ratio in Nov 2014. \newline
Cases 11-13 have Ly-$\alpha$/Ly-$\beta =$ 2.6, 3.2 and 3.3 respectively, suggesting that a process with a small line ratio has contributed significantly to the Ly-$\beta$ brightness. Electron impact on other neutral species cannot be a strong source of OI1027 in these cases, as they would show stronger emissions in other lines (e.g. OI1356 for \ce{O2} and \ce{CO2}) which were not observed. Resonant scattering of solar flux on atomic hydrogen would generate Ly-$\alpha$ emissions 300 times brighter than Ly-$\beta$, greatly increasing the line ratio. It is also very unlikely at the low cometocentric distances ($\sim 10$~km) as neutral molecules are advected away from the nucleus \citep[timescale of the order of seconds;][]{Galand2016} before they can undergo photodissociation \citep[timescale $> 10^5$ s;][]{Huebner2015}. \newline
Prompt-photodissociation, which gives a Ly-$\alpha$/Ly-$\beta$ ratio of 1.4 \citep{Hans2015}, could generate the lower line ratio. The emission of OI1304 from this process is also very weak, as a spin forbidden transition would be required \citep{Wu1988}. However, if we had significantly underestimated the efficiency of this process in the model, a discrepancy would be seen in the other cases. A large increase in the column of water in cases 11-13 (by a factor of 8, 15 and 5, respectively) would bring the modelled Ly-$\beta$ into agreement with the observed brightnesses, whilst generating few emissions in the other atomic lines due to the low emission frequencies (dark blue, Figs. \ref{fig: emission freq}a~\&~b). However, this would result in a neutral column comprising 80\% water in cases 11 and 12 (50\% in case 13), which is inconsistent with the concurrent mass spectrometer measurements (see Fig \ref{fig: column density}).The required mixing ratio would also be incongruous with wider outgassing trends, as the outgassing rate of \ce{CO2} in the southern hemisphere was larger than that of \ce{H2O} in Mar 2016 (Case 11) and was dominant over the outgassing of \ce{H2O} near the end of mission \citep[Cases 12 and 13; ][]{Gasc2017MNRAS, Laeuter2018}. The source of the Lyman-$\beta$ emissions in cases 11-13 remains unclear.}}
\section{Corotating interaction regions in summer 2016}\label{sec: CIRs}
\subsection{Selection of cases}\label{sec: CIR selection}
Having confirmed that the impact of suprathermal electrons on cometary neutrals is a major source of emissions during quiet periods in the southern hemisphere in Section \ref{sec: Nadir}, we apply the multi-instrument analysis to the CIRs observed in the summer of 2016. The CIR was observed over four solar rotations throughout the summer of 2016 \citep{Hajra2018}, from the start of June until the beginning of September. In the present study, we consider two occurrences: one over the 9 and 10 July, and one on the 4 August. \textcolor{edits}{Enhanced FUV emissions were observed on two additional solar rotations in early and late September. However, observations in the FUV occurred infrequently during these events, so we could not study the temporal variability of the emissions.}  \newline
During these events, the suprathermal electron flux can vary by a factor of 100 within several hours, as seen during the first periods on 4 Aug and 9 July (see Tables \ref{tab: August column} and \ref{tab: July column} \textcolor{comments}{and Fig. \ref{fig: Aug Electron Spectra}}). The brightness of the FUV emission lines are also seen to vary by a factor of ten within the same time periods (see Figs.~\ref{fig: August} and \ref{fig: July}). This scale of variation in the emission brightness has also been observed by \cite{Galand2020} during a solar event in October 2014 and \cite{Noonan2018} during a CME in summer 2015. To capture the variability, we use the highest temporal resolution available for both the measurements of the electron flux and the observed FUV brightness (10~\si{\minute}). The emission frequencies are calculated from individual electron flux measurements, once the flux has been corrected for the spacecraft potential (see Section \ref{sec: electron flux}). The modelled brightness is plotted in Figures \ref{fig: August} and \ref{fig: July} at the time resolution of the electron spectrometer. We do not co-add consecutive spectra from the FUV spectrograph, as in Section \ref{sec: Alice Obs}, but instead evaluate the brightness at three distinct regions of the viewing slit for each spectrum. The black vertical lines are therefore indicative of the spatial variability of the emissions along the Alice slit. In Figures \ref{fig: August} and \ref{fig: July}, the three regions associated with rows 8-11, 13-16, and 18-21 are given by the red, black, and blue crosses, respectively. The measurements taken simultaneously in the 3 regions of the FUV spectrograph slit are linked by vertical black lines. \newline
Neither the case in July nor the one in August 2016 had a nadir viewing geometry as was used in Section \ref{sec: Nadir}. On 4 August 2016, the FUV spectrograph was viewing off the limb of the nucleus, whereas on the 9-10 July 2016 the spectrograph was pointed at the nucleus but off-nadir. In both situations it is not possible to derive the column density along the line of sight from the ROSINA in-situ measurements as was done in cases 1-8 in Section \ref{sec: nadir 1-8}. \textcolor{comments}{{Variation across the nucleus' surface in both the density and composition of the outgassing neutrals means we cannot extrapolate from in situ measurements to calculate the column density along the line of sight.}} For the August case, the column densities of \ce{CO2} and \ce{H2O} are derived from remote observations by VIRTIS-H and MIRO spectrometers, respectively, (see Section \ref{sec: Neutral Column}) over each of the intervals of FUV observations. \newline
In the July event, the surface of the nucleus was illuminated, which prevents the estimation of reliable column densities using either of the spectrometers. We, therefore, set the column density of \ce{CO2} during each interval such that the scale of the modelled and observed brightnesses of the OI1356 within each interval agree, which is consistent with the approach in cases 1-8 in Section \ref{sec: nadir 1-8}. \newline
The column density, taken constant with each interval during the CIR events considered are given in Tables \ref{tab: August column} and \ref{tab: July column}. The different periods, with distinct neutral column densities, are indicated by different coloured lines in Figures \ref{fig: August} and \ref{fig: July}. During each interval the nucleus moved slowly within the field of view of the spectrograph, justifying the constant column density during each interval. However, between these periods the spacecraft underwent manoeuvres in which the region of the coma observed changed rapidly, which justifies the variation of the column density from one interval to the next. As the column density is taken as constant in each interval, the variation of the modelled brightness during the time periods is only due to the fluctuations of the measured suprathermal electron flux.
The average electron fluxes in each period of the FUV spectra are shown in Figures \ref{fig: Aug Electron Spectra} and \ref{fig: Jul Electron Spectra} for August and July, respectively. The electron flux in period 1 of the August case is split into two parts, comprising the peaks (dark blue) and troughs (light blue) of the electron flux as seen in the modelled brightness in Fig.\ref{fig: August} (dark blue). 
\subsection{4 August 2016} \label{sec: CIR Aug}
\subsubsection{Overview}
\begin{figure}[htbp]
    \includegraphics[width = 0.5 \textwidth]{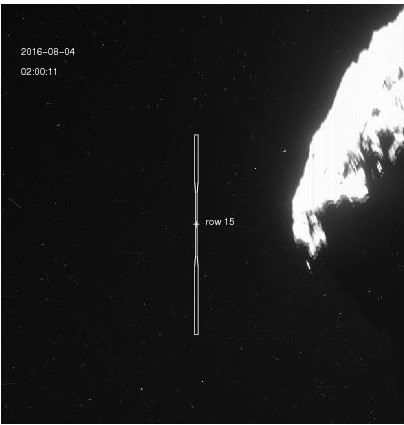}
    \caption{\textcolor{comments}{{OSIRIS WAC image \citep{keller2007osiris} from 4 Aug 2016 02:00~UT during period 1 (see Section \ref{sec: Aug P1}) with the Alice viewing slit shown in white. The line of sight passes within $\sim500$~m of the comet nucleus.}}}
    \label{fig: Aug viewing Geom}
\end{figure}
\begin{figure*}[htbp]
    \begin{tikzpicture}[node distance= 1cm, baseline=(current  bounding  box.center)]
    \node (aug) at (0,0) {\includegraphics[width = \textwidth]{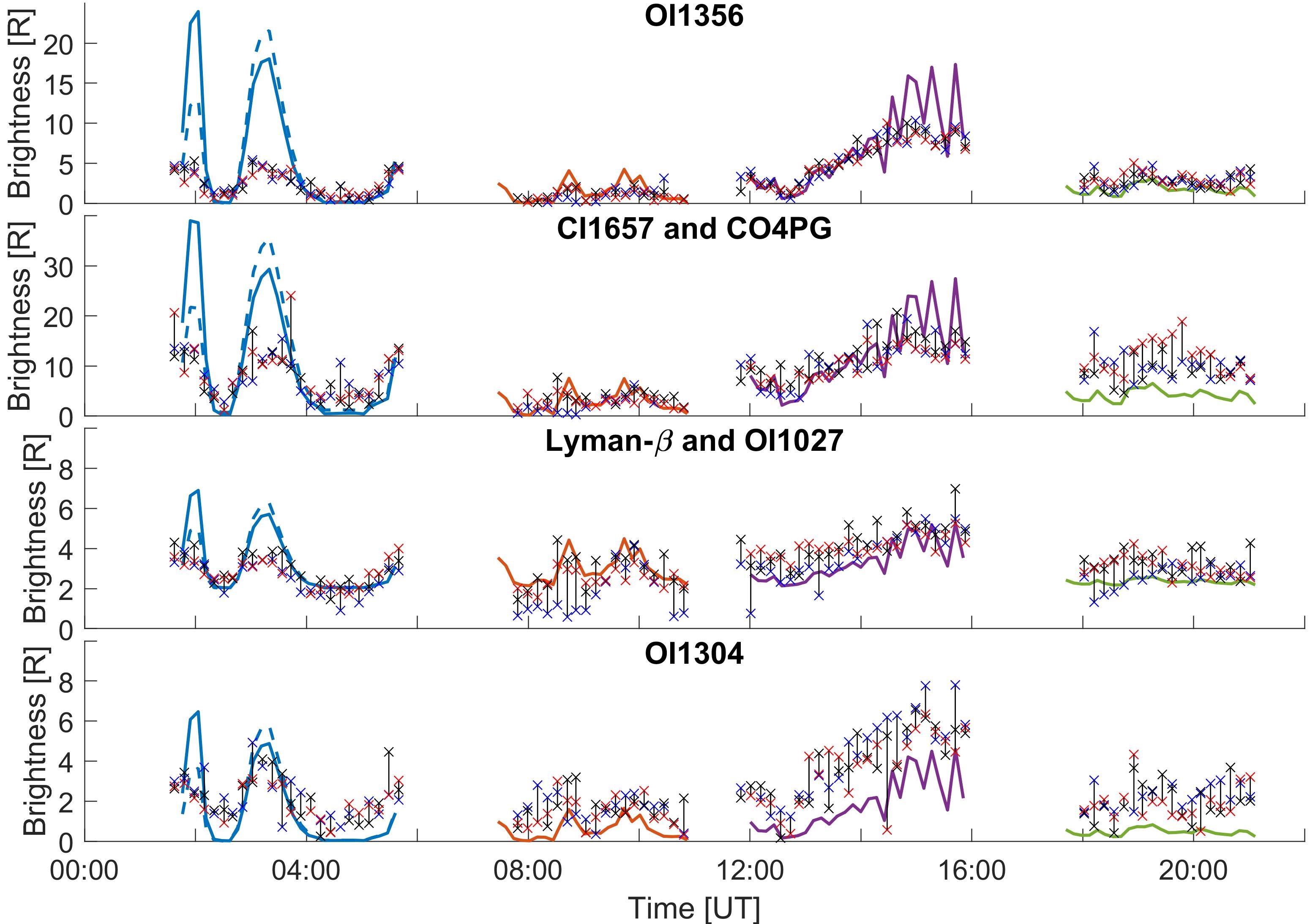}};
    \node [above left = 12.2 and 7.2 of aug.base, font =\large] (A) {(a)};
    \node [above left = 9.0 and 7.2  of aug.base, font =\large] (B) {(b)};
    \node [above left = 6.2 and 7.2 of aug.base, font =\large] (C) {(c)};
    \node [above left = 3 and 7.2 of aug.base, font =\large] (D) {(d)};
    \end{tikzpicture}
    \caption{Comparison of the observed (crosses) and modelled (coloured lines) brightness of four emission features, OI1356 (a), CI1657 and CO4PG (b), Lyman-$\beta$ and OI1027 (c) and OI1304 (d), during four time periods, during which a CIR was observed at comet 67P, on 4 Aug 2016 (see Table \ref{tab: August column}). The observed brightnesses are from three regions of the FUV spectrograph slit (Rows 8-11 [red], 13-16 [black] and 18-21 [blue]). Vertical black lines link simultaneous measurements of the brightness.}
    \label{fig: August}
\end{figure*}
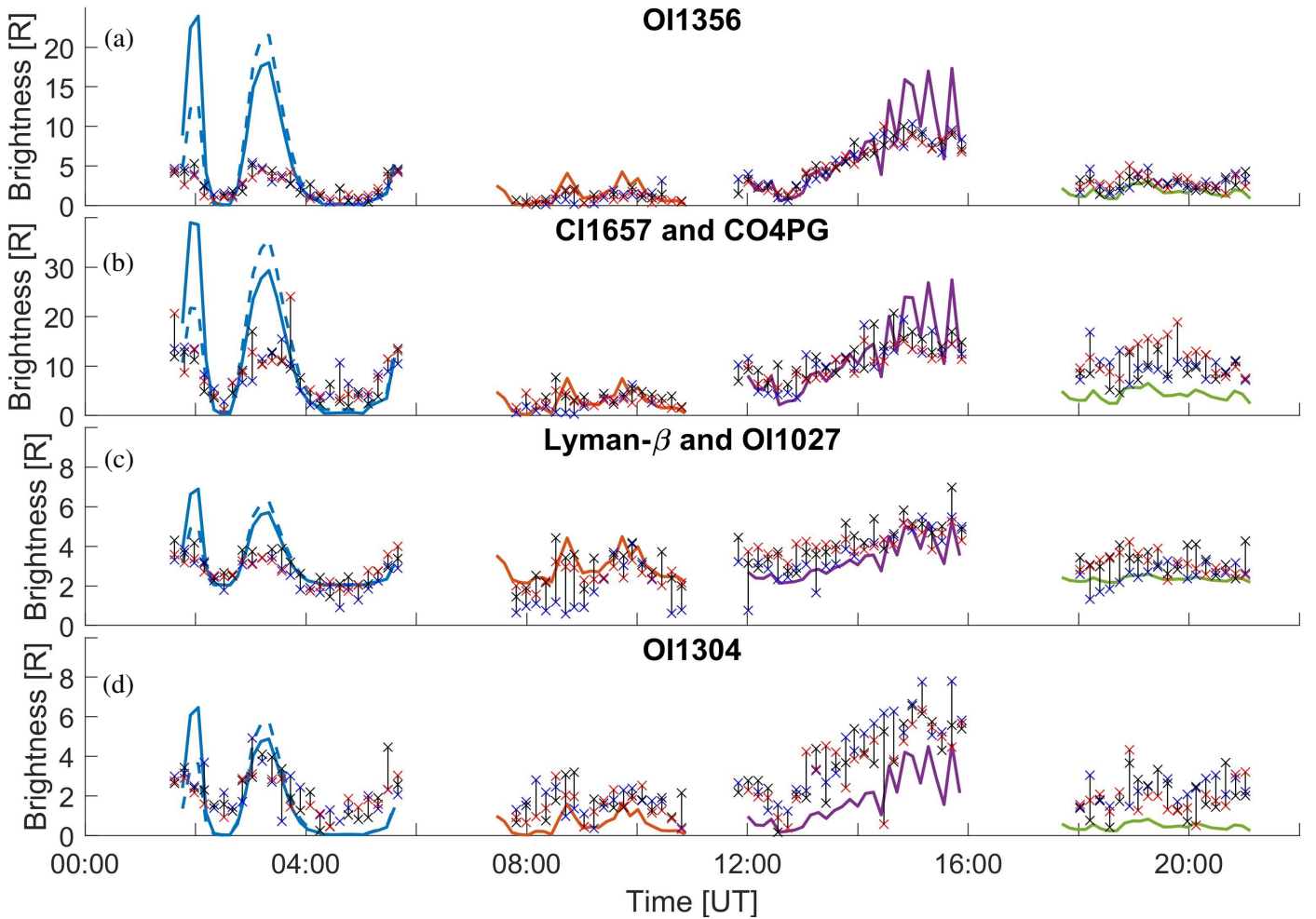

On this day the viewing geometry was limb \textcolor{comments}{{(Fig \ref{fig: Aug viewing Geom})}}, so \textcolor{edits}{we have used column density measurements from the VIRTIS-H and MIRO spectrometers} (see Section \ref{sec: Neutral Column}). The column density measurements taken in the four time periods are outlined in Table \ref{tab: July column}. This solar event occurred near the end of mission at 3.5~\si{\astronomicalunit}, when \ce{CO2} was dominant over \ce{H2O} in the southern hemisphere as seen in Table \ref{tab: August column}, which is consistent with a volume mixing ratio of $\ce{H2O}/\ce{CO2}\sim 0.05$ from \cite{Laeuter2018}. During period 2, the line of sight of the spectrograph passed over the neck region of the comet, which had an increased abundance of water \citep{Migliorini2016}, resulting in the enhanced water column density. \ce{CO} was not detected during any of the four periods and the 3$\sigma$ upper-limits of the column density derived from the sub-mm spectrometer are given in Table \ref{tab: August column}. \textcolor{comments}{{The FUV emission spectra display weak CO band emissions which indicates there is some CO in the coma, although insufficient to be detected by MIRO.}} In Figure \ref{fig: August}, we have assumed that there is no \ce{CO} present in the coma as we do not have a good constraint on the CO mixing ratio throughout the column. The spectrometers could not measure the column density of \ce{O2}, so we have assumed there is no \ce{O2} present in the column along the line of sight. Towards the end of mission, the mixing ratio $\ce{O2}/\ce{CO2}$ decreased \citep[$<1$\%,][]{Laeuter2018} and \ce{O2} was less abundant in the southern hemisphere \citep{Hassig2015}, so we would not expect significant emissions from \ce{e + O2}. \textcolor{edits}{Including a few percent of \ce{O2} relative to the \ce{H2O} column density has negligible impact on the emission brightness, as emissions of the atomic oxygen lines are dominated by electron impact on \ce{CO2}.} \newline
\begin{table*}[htbp]
    \caption{Properties of the four intervals on 4 Aug 2016 at 3.5~\si{\astronomicalunit}. \textcolor{comments}{{The first period has been split into the peaks and troughs in the electron flux, which can be seen in Fig.~\ref{fig: August}. These are the same periods as shown in Fig. \ref{fig: Aug Electron Spectra} in dark blue (peaks) and light blue (troughs).}} \tablefoottext{1}{From VIRTIS-H.}\tablefoottext{2}{From MIRO.}\tablefoottext{3}{3$\sigma$ upper limits from MIRO. }}
    \label{tab: August column}
    \centering
    {\tabulinesep = 1.2mm 
    \begin{tabular}{||>{\centering\arraybackslash}m{1.7cm}|>{\centering\arraybackslash}m{1.5cm}|>{\centering\arraybackslash}m{1.5cm}|>{\centering\arraybackslash}m{2cm}|>{\centering\arraybackslash}m{2cm}|>{\centering\arraybackslash}m{2cm}|>{\centering\arraybackslash}m{2.8cm}||}
    \hline
         \thead{Period} & \thead{Start Time \\ {[UTC]} } & \thead{End Time \\ {[UTC]} } & \thead{\ce{CO2} Column \\ Density\tablefootmark{1} \\ {[$10^{14}$~\si{\per\square\centi\metre}]} } & \thead{\ce{H2O} Column \\ Density\tablefootmark{2} \\{[$10^{14}$~\si{\per\square\centi\metre}]}}  &  \thead{\ce{CO} Column \\ Density\tablefootmark{3} \\ {[$10^{14}$~\si{\per\square\centi\metre}]} } & \thead{Ave. Electron Flux \\ for 20 - 60 \si{\electronvolt} \\ {[$10^7$ ~\si{\per\square\centi\metre\per\second\per\electronvolt}]}} \\
         \hline
         1 - Peaks & \multirow{2}{*}[-0.3em]{01:38:59} & \multirow{2}{*}[-0.1em]{05:42:30} & \multirow{2}{*}[-0.1em]{$6.21\pm 0.16$} & \multirow{2}{*}[-0.1em]{$0.5\pm0.03$} & \multirow{2}{*}[-0.1em]{<1.1} & 15.5 - 44.9 \\
         1 - Troughs & & & & & & 0.56 - 10.1\\
         2 & 07:25:24 & 10:59:15 & $2.65\pm0.17$ & $2.24\pm 0.09$ & -- & 0.77 - 23.3 \\ 
         3 & 11:51:51 & 15:56:18 & $8.43\pm0.14$ & $0.36\pm 0.02$ & <1.3&3.16 - 20.4\\ 
         4 & 17:38:16 & 21:12:03 & $7.93\pm0.23$ & $0.31\pm0.02$ & -- & 3.48 - 7.95\\ 
         \hline
    \end{tabular}
    }
\end{table*}
\begin{figure}
    \includegraphics[width=0.5\textwidth]{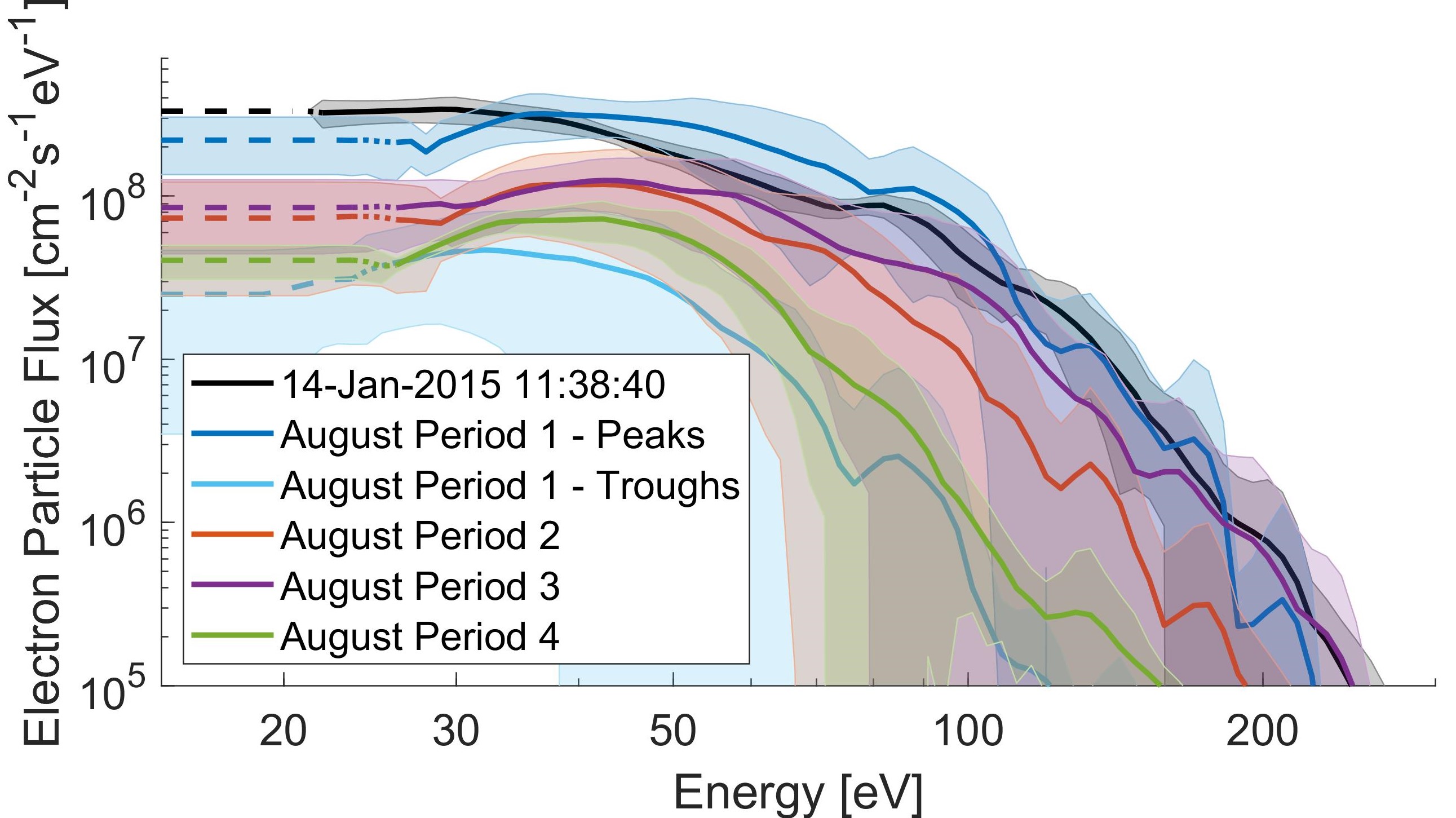}
    \caption{\textcolor{comments}{{Average electron particle flux during the three periods listed in Table \ref{tab: August column}, with the standard deviation of the flux shown by the shaded regions. The colour of each period is the same as in Fig. \ref{fig: August}. The first period has been split into the peaks (dark blue) and troughs (light blue) in the electron flux to highlight the variability in this period. The dotted lines indicate energies where some of the RPC/IES measurements during the period could not probe, due to a high spacecraft potential (see Section \ref{sec: electron flux}). The dashed lines denote energies that could not be probed by any of the RPC/IES scans during the period. The electron flux from Case 1 in the nadir study (see Section \ref{sec: Nadir} \& Fig. \ref{fig: electron flux}) is plotted for comparison (black).}}}
    \label{fig: Aug Electron Spectra}
\end{figure}
During the CIR, the variation in the suprathermal electron flux is mirrored in the brightness of the emissions in all four of the selected lines: OI1356 (Fig.~\ref{fig: August}a), CI1657 and CO4PG (Fig.~\ref{fig: August}b), Ly-$\beta$ and OI1027 (Fig.~\ref{fig: August}c), and OI1304 (Fig.~\ref{fig: August}d). This suggests that all of these emissions are strongly driven by electron impact on cometary neutrals. For limb viewing, the emissions originate from a distant region of the coma, which cannot be probed in situ. The observed correlation between the in situ measurements of the electron flux and the remote measurements of the FUV spectrograph suggest that the variations in the electron flux occur on large scales.
\textcolor{comments}{{The electron fluxes during each period (see Fig.~\ref{fig: Aug Electron Spectra}) peak at 40-50~eV, which is not seen in the electron spectra during the nadir cases in quiet periods (see Fig. \ref{fig: electron flux}). This may result from the higher density and more energetic solar wind electrons entering the coma that are associated with the CIR.}}
\subsubsection{Periods 2-4}\label{sec: Aug P2-4}
The modelled brightness of the OI1356 emissions closely replicate both the variations and the magnitude of the observed emissions in periods 2-4 (see Fig.~\ref{fig: August}a). The emissions are dominated by \ce{e + CO2}, which accounts for 98\% of the modelled emissions. This is consistent with our findings from the nadir cases over the southern hemisphere (see Fig.~\ref{fig: Nadir Quiet}a) for which the neutral composition was derived from the mass spectrometer (see Section \ref{sec: nadir 1-8}). The close agreement in periods 2-4 suggests that there are no other significant sources of OI1356 in these cases. The lower OI1356 brightness during the second time period is a result of the lower column of \ce{CO2} compared to the following period (see Table \ref{tab: August column}). In period 3, the OI1356 emission frequency from \ce{e + CO2} increases from $7.1\times 10^{-10}$~\si{\per\second} at 12:35~UT to $2.1\times 10^{-8}$~\si{\per\second} at 15:42UT, which causes the concurrent rise in the OI1356 brightness. \newline
\textcolor{comments}{{Period 4 has fewer emissions due to the small electron flux (green, Fig. \ref{fig: Aug Electron Spectra} \& Table \ref{tab: August column}), which results in a lower emission frequency. In period 3 the emission frequency of OI1356 from \ce{e + CO2} reached $2.1\times10^{-8}$~s$^{-1}$, five times the peak frequency in period 4 ($3.9\times10^{-9}$~s$^{-1}$).}} \newline
When looking nadir in Section \ref{sec: nadir cases}, the CI1657 emissions were dominated by electron impact on \ce{CO2}, with only a small contribution from fluorescence of CO (dark green, Fig.~\ref{fig: Nadir Quiet}d) when the suprathermal electron flux was high. For limb viewing, the modelled brightness of CI1657, driven only by electron impact on \ce{CO2} (without any contribution from fluorescence of CO),  well reproduces the observed brightness throughout the periods 2 and 3 (see Fig.~\ref{fig: August}b), which is consistent with there being no significant column of \ce{CO} present along the line of sight ($\ce{CO}/\ce{CO2}< 0.05$). Furthermore, we would not expect any significant contribution from fluorescence in this case, as \ce{e + CO} is a more efficient source of emissions when the suprathermal electron flux is large (see cases 1-8, Fig.~\ref{fig: emission freq}). \textcolor{comments}{{In period 4, a slight underestimation of CI1657 by the model may result from the lack of CO column included in the model. A $\ce{CO}/\ce{CO2}$ ratio of 0.2 would explain the disparity, which is only slightly larger than the ratio in production rates ($\ce{CO}/\ce{CO2}=0.1$) in the southern hemisphere at this time \citep{Laeuter2018}.}}\newline
For limb viewing, emissions from the IPM contribute to the observed Ly-$\beta$ brightness, unlike in the nadir case (see Section \ref{sec: Alice Obs}). As such we include a 2-rayleigh background contribution from the IPM in this line. The Ly-$\beta$ brightness exhibits the same variations as measured in the electron flux throughout the CIR (see Fig.~\ref{fig: August}c), suggesting that dissociative excitation by electron impact is a significant source of this line. \newline
\textcolor{comments}{{The brightness of Ly-$\beta$ emissions is well captured throughout periods 2-4, which suggests there are no other major sources of emissions.}}  The contribution from prompt-photodissociation of \ce{H2O} to the total brightness is negligible throughout the CIR.  In the limb viewing geometry, the extended coma is also observed, in which photodissociation is a significant source of neutral atoms. \textcolor{comments}{{There could be some contribution from resonant scattering off atomic hydrogen in the extended coma, but the agreement between the modelled and observed brightnesses in Fig.~\ref{fig: August}b suggests there are not significant emissions from this source. However, it is difficult to estimate it independently as there is no constraint on the column of atomic hydrogen from remote instrumentation on Rosetta.}} \newline
Again, the fluctuations of the OI1304 emission brightness mirror the changes in the suprathermal electron flux (see Fig.~\ref{fig: August}d), indicating that \ce{e + X} is a major source of emission. In the nadir cases, these emissions were driven by electron impact on both \ce{CO2} and \ce{H2O} (see Fig.~\ref{fig: Nadir Quiet}b) and when the suprathermal electron flux was low there were no emissions in this line (see cases 10-14, Fig.~\ref{fig: Nadir Quiet}). \textcolor{comments}{{When looking at the limb, we observe substantial emissions ($\sim\!3$~R) of OI1304 when the average electron flux from $20-60$~\si{\electronvolt} is less than $10^7$~\si{\per\square\centi\metre\per\second\per\electronvolt}. The disparity between the observed and modelled emissions in Fig.~\ref{fig: August}d is approximately constant throughout periods 2-4, suggesting that the residual emissions are not driven by electrons which showed strong fluctuations over the same time period. Similar to Ly-$\beta$, there may be a contribution from resonant scattering from atomic oxygen along the line of sight, but with a lack of constraints we cannot determine whether this is sufficient to explain the discrepancy.}}
\subsubsection{Period 1}\label{sec: Aug P1}
\textcolor{comments}{{During the first period on 4 Aug 2016, the variations in the electron flux with time are mirrored by the changes in the line brightnesses for all four atomic lines. There are two strong peaks in the electron flux (dark blue, Fig. \ref{fig: Aug Electron Spectra}) from 01:45-02:20~UT and 02:40-03:40~UT during which all four of the atomic lines considered also exhibit peaks in the brightnesses. From 03:40~UT until 05:20~UT, the electron flux is small (light blue, Fig. \ref{fig: Aug Electron Spectra}) and no significant FUV emissions are observed in the atomic lines. The 2~rayleighs of Ly-$\beta$ emissions observed during this trough are attributed wholly to interplanetary emissions that are seen with a limb viewing geometry.  An increase in the electron flux at the end of the first period (after 05:20~UT) also coincides with a rise in the brightnesses of the atomic lines. \newline
However, the scale of the fluctuations related to the two strong peaks in the modelled emissions greatly exceeds those observed with the FUV spectrograph. This disparity in scale could be attributed to either a poorly estimated column density or in the electron flux along the line of sight. 
A reduction of the \ce{CO2} column, which was the dominant emission source in period 1, by a factor of 0.4 gives a close agreement between the modelled and observed brightnesses for all four of the atomic lines. With this reduction, there is a slight underestimation of the OI1304 brightness, which is consistent with the results in the later periods. Variation of the \ce{CO2} column density within the first period could drive the difference in scale, so we have considered higher time resolution VIRTIS data, which splits the first period into four parts. A slightly lower \ce{CO2} column is found during the first peak ($3.37\times10^{14}$~cm$^{-2}$, 01:40-02:40~UT), but during the second peak the \ce{CO2} column density is found to be larger ($7.4\times10^{14}$ cm$^{-2}$, 02:40-03:40~UT). The result of using these column is plotted with a dashed blue line in Figure \ref{fig: August}. \newline
Alternatively, the assumption that the suprathermal electron flux is constant along the line of sight may break down when viewing the extended coma. The electron flux is measured at a cometocentric distance of 12~km, but the limb observed passes within roughly 500~m of the nucleus surface at the closest point (see Fig.~\ref{fig: Aug viewing Geom}) so there could be some variation along the line of sight. Using the higher time resolution \ce{CO2} columns from VIRTIS, the electron flux would have to be reduced by a factor 0.6 during the first peak and 0.3 during the second peak to reach a close agreement between the observed and modelled brightnesses. 
However, if this assumption were invalid, a similar disparity to that seen in period 1 would be expected in periods 2-4, which is not the case. The viewing geometry has been compared between period 1 and periods 2-4 and there is no obvious distinction (all four periods have a line of sight passing within 1~km of the nucleus) that would cause a difference in the electron behaviour. \newline
As the temporal variations of the electron flux and the FUV emissions are well correlated for all four atomic lines, these emissions are driven by electron impact and generated by the variations in the CIR. However, it is not clear why the brightness of the emission lines during the two large peaks in period 1 are overestimated in the model.}}
\subsection{9 - 10 July 2016} \label{sec: CIR Jul}
\begin{figure*}[htbp]
    \begin{tikzpicture}[node distance= 1cm, baseline=(current  bounding  box.center)]
    \node (jul) at (0,0) {\includegraphics[width = \textwidth]{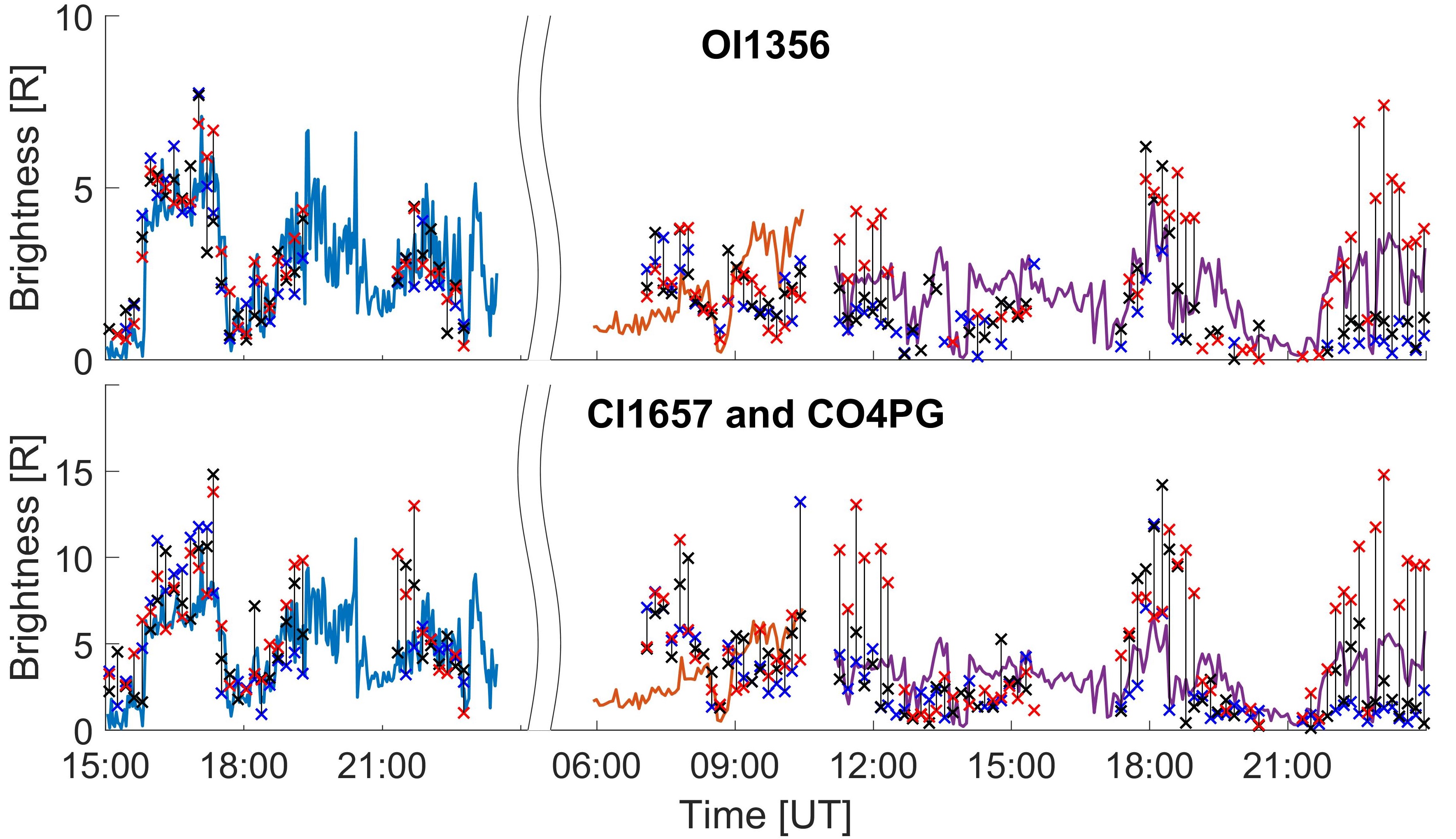}};
    \node [above left = 9.3 and 7 of jul.base, font =\large] (A) {(a)};
    \node [above left = 4.7 and 7 of jul.base, font =\large] (B) {(b)};
    \node [above left = 9.3 and 4.5 of jul.base, font = \large] (Jul9){July 9th};
    \node [above right = 9.3 and 3 of jul.base, font = \large] (Jul10){July 10th};
    \end{tikzpicture}
    \caption{Comparison of the observed and modelled brightness of the OI1356 (a) and CI1657 emission lines from 9 July 15:00~UT to 11 July 00:00~UT 2016. The plots follow the same format as outlined in Figure \ref{fig: August}. The modelled brightnesses shown here are only driven by electron impact on \ce{CO2}.}
    \label{fig: July}
\end{figure*}
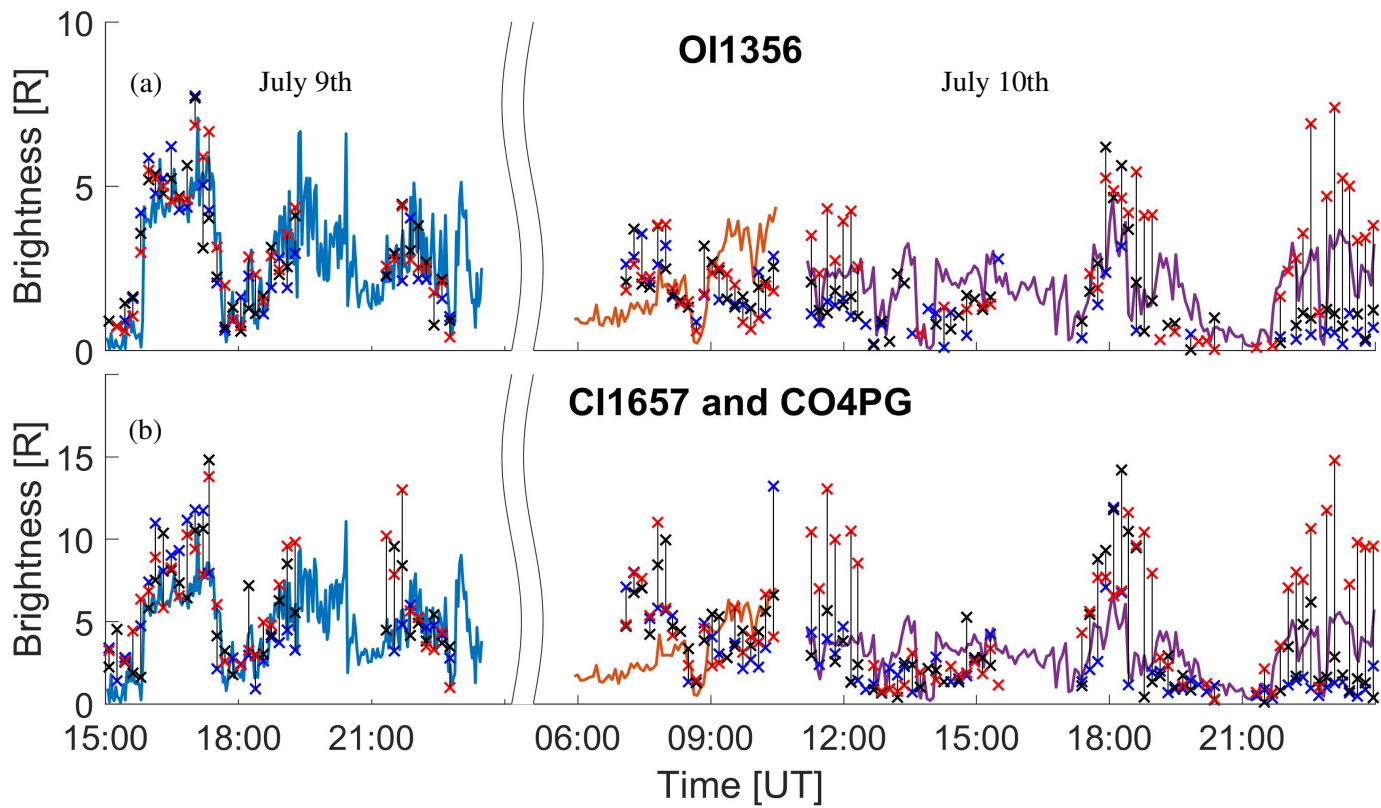

\textcolor{comments}{{During the CIR on 9-10 July 2016, the FUV spectrograph is viewing the illuminated nucleus, which pollutes the FUV emission spectra at long wavelengths ($>1500$~{\AA}) and prevents reliable measurements of the column density by either spectrometer. There are only a few periods when the emissions driven by a CIR were observed by Alice, so despite the difficulties in the analysis the reflected solar photons present, this period is still of great interest.}} \newline
\textcolor{comments}{{In situ measurements do not provide a good constraint on the density and composition of the neutral gas column due to the off nadir view.}} Therefore, we adjust the column density during each of the periods to match the observed OI1356 emission brightness, which is consistent with our findings of \ce{CO2} driving this line over the southern hemisphere (Sections \ref{sec: nadir 1-8} and \ref{sec: CIR Aug}). The resulting column densities are given in Table \ref{tab: July column}, under the assumption of a pure \ce{CO2} coma.  \textcolor{comments}{{The adjusted column densities of \ce{CO2} are between 0.88 and 1.6 times the radial column density derived from in situ measurements of the neutral density (see Sec. \ref{tab: July column}). This is a good agreement given the variation in the outgassing across the surface and the longer path through the coma due to the viewing geometry. The off-nadir angle during periods 1-3 was roughly 10$^\circ$, so the spatially variable outgassing is likely the major driver of the difference in column density.}} The in situ measurements used for this comparison have been corrected assuming the neutral gas is only \ce{CO2} \citep{Gasc2017MNRAS}. \newline
In Figure \ref{fig: July}, we plot only the modelled and observed brightness of the OI1356 (a) and CI1657 (b) lines, as we have no constraint on the \ce{H2O} column density. The Ly-$\beta$ and OI1304 emissions both have significant contributions from \ce{e + H2O}, which we cannot constrain with the OI1356 emissions (see Section \ref{sec: nadir 1-8}).\newline
\textcolor{comments}{{Reflection of solar flux off the illuminated nucleus is seen in the Alice spectra at long wavelengths, strongly contributing to the atomic line at 1657~{\AA}. In order to distinguish the CI1657 emissions from the coma from those reflected off the nucleus, we subtract the solar spectrum from the Alice observations during this event as outlined in Appendix \ref{sec: Alice analysis}. OI1356 is a weak line in the solar spectrum due to the associated forbidden transition, so is unaffected by this correction. Reflected solar flux would also contribute to the Ly-$\beta$ and OI1304 brightnesses, but these lines are not considered in this section.}} \newline
\begin{table*}[htbp]
    \caption{Properties of the three intervals on 9-10 July.}
    \label{tab: July column}
    {\tabulinesep = 1.2mm 
    \begin{tabular}{||>{\centering\arraybackslash}m{1cm}|>{\centering\arraybackslash}m{2.5cm}|>{\centering\arraybackslash}m{2.5cm}|>{\centering\arraybackslash}m{2cm}|>{\centering\arraybackslash}m{2.3cm}|>{\centering\arraybackslash}m{2.8cm}||}
    \hline
         \thead{Period} & \thead{Start Time \\ {[After 9 July} \\ { 2016 00:00 UTC]} } & \thead{End Time \\ {[After 9 July} \\ {2016 00:00 UTC]}} & \thead{\ce{CO2} Column\\ Density \\{[$10^{14}$~\si{\per\square\centi\metre}]}} & \thead{Total Radial \\ Column Density \\ {[$10^{14}$~\si{\per\square\centi\metre}]}} & \thead{Ave. Electron Flux \\ for 20 - 60 \si{\electronvolt}\\ {[$10^7$ ~\si{\per\square\centi\metre\per\second\per\electronvolt}]}}\\
         \hline
         1 & 15:00:00 & 23:30:00 & \textcolor{comments}{6.0} & 3.83 & 0.16 - 15.3 \\
         2 & 29:52:00 & 34:30:00 & \textcolor{comments}{2.0} & 2.25 & 2.46 - 25.0 \\ 
         3 & 35:10:00 & 48:00:00 & \textcolor{comments}{1.2} & 0.76 & 1.97 - 36.2\\
         \hline
    \end{tabular}
    }

\end{table*}
\begin{figure}
    \includegraphics[width=0.5\textwidth]{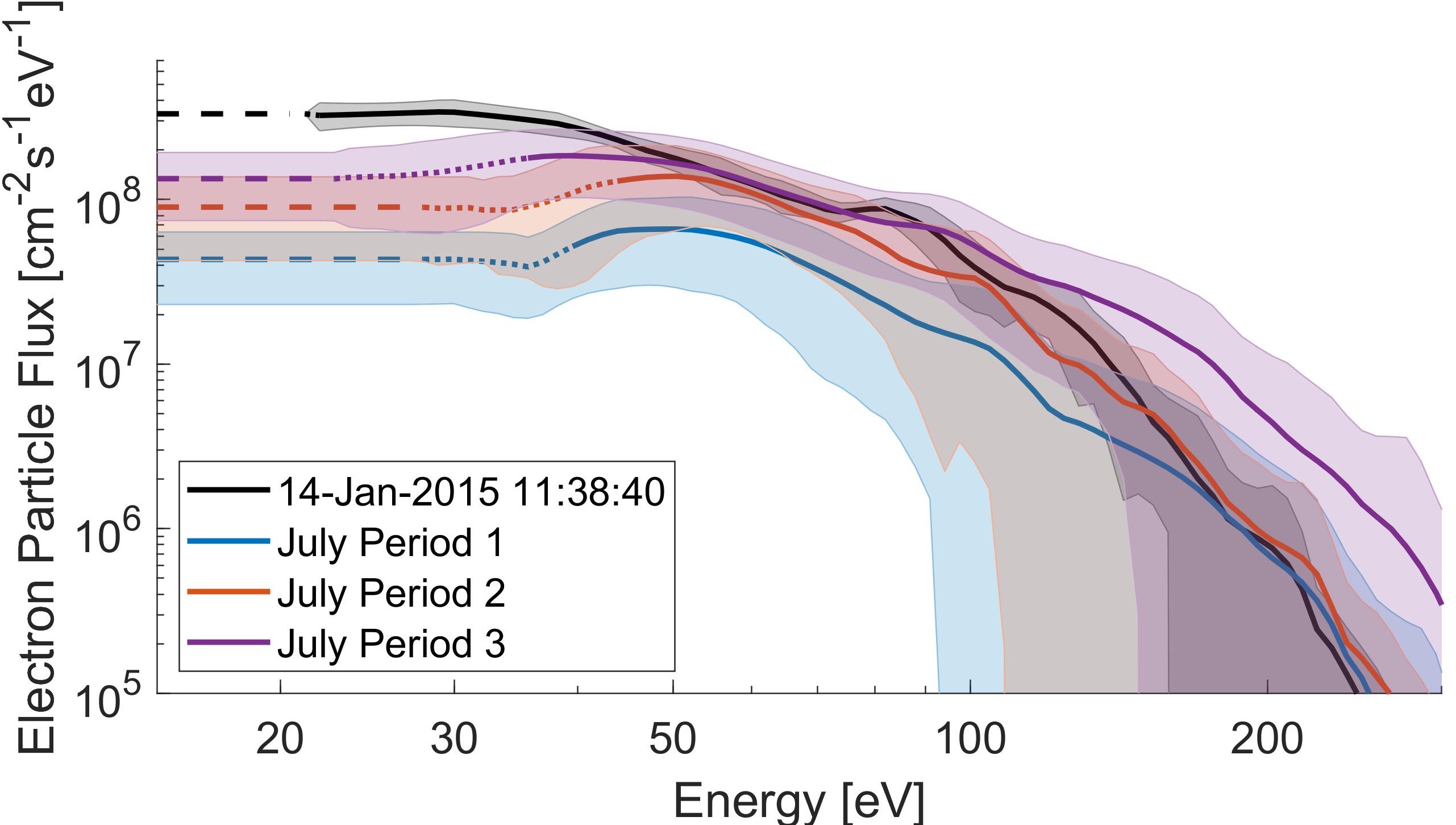}
    \caption{\textcolor{comments}{{Average electron particle flux during the three periods listed in Table \ref{tab: July column}, with the standard deviation of the flux shown by the shaded regions. The colour of each period is the same as in Fig. \ref{fig: July}. The format of the plot is the same as outlined in Figure \ref{fig: Aug Electron Spectra}.}}}
    \label{fig: Jul Electron Spectra}
\end{figure}
The OI1356 brightness closely matches the variation of the electron flux during this event. Between 16:00 and 18:00~UT  on 9 July, the emission frequency of OI1356 from \ce{CO2} increases from $10^{-9}$~\si{\per\second} at 15:45~UT, plateaus at $9\times10^{-9}$~\si{\per\second} until 17:30~UT, and then drops to $4\times10^{-10}$~\si{\per\second} at 17:45~UT. 
The close agreement in the fine structure can be seen between 07:30 and 09:30UT on 10 July. The emission frequency of OI1356 from \ce{e + CO2} increases from $5.7\times10^{-9}$~\si{\per\second} to a maximum $1.3\times10^{-8}$~\si{\per\second} at 08:00~UT, followed by a decrease to $1.1\times10^{-9}$~\si{\per\second} at 08:40~UT. At 08:50~UT, the emission frequency and brightness of the OI1356 line both increase rapidly to $2.0\times10^{-8}$~\si{\per\second} and 3.2~R, respectively, before plateauing. \newline
\textcolor{comments}{{The observed CI1657 emissions display the same structures seen in the OI1356 brightness (e.g. between 16:00 and 18:00~UT on 9 July; Fig.~\ref{fig: July}b), capturing both the magnitude and variability of the fluctuations. On 10 July, a peak in the observed CI1657 brightness at 18:00~UT is mirrored by an increase in the CI1657 emission frequency to $5.59\times10^{-8}$~s$^{-1}$, before a decrease in both the observed emissions from the coma and the electron flux until 21:00~UT. When there are few suprathermal electrons around 21:00~UT 10 July, there are almost no observed emissions from the coma, demonstrating that there are no other major sources of emissions from the coma during this event. The good agreement between the in situ electron measurement and remote FUV observations strengthens the case that the electron variations occur over large scales.}} \newline
\section{Conclusion}\label{sec: Conclusion}
The Rosetta spacecraft's proximity to the nucleus of 67P, throughout the two-year escort phase, provided us with the opportunity to observe FUV emissions from within the coma. We have performed the first forward modelling of FUV emissions in the southern hemisphere of comet 67P using an extension of the multi-instrument analysis of \cite{Galand2020} and introducing carbon lines for the first time. When observing the shadowed nucleus at large heliocentric distances, the analysis we have applied well reproduces the observed brightnesses of the selected FUV emission lines (OI1356, OI1304, Ly-$\beta$ and OI1027, CI1657 and CO4PG, and CII1335) for periods with large suprathermal electron fluxes (see Section \ref{sec: nadir 1-8}) . Therefore, dissociative excitation by electron impact is a key source of the FUV emissions in the southern hemisphere away from perihelion. When the suprathermal electron flux was small (Averaged electron flux $<2\times10^7$~\si{\per\square\centi\metre\per\second\per\electronvolt} between 20 and 60 \si{\electronvolt}), no emission of either of the OI lines or the CII line \textcolor{edits}{($B^{\text{OI1304}}<0.55$R, $B^{\text{OI1356}}<0.79$R, $B^{\text{CII1335}}<0.22$R)} were observed, indicating that energetic electrons in the coma are the dominant driver of these emissions (see Section \ref{sec: nadir 9-13}). \textcolor{comments}{{In the low flux cases, fluorescence of CO was a larger source of emissions at 1657~{\AA} compared to many of the cases with a high electron flux (see Fig.~\ref{fig: Nadir Quiet}d), due to the larger total column density (see Table~\ref{tab: Nadir cases}). The relative contribution of CO fluorescence in the low flux cases was also enhanced by the small emission frequency from electron impact processes (see Fig.~\ref{fig: emission freq}d), which had dominated in the high flux cases (cases 1-8, Fig~\ref{fig: Nadir Quiet}d).}} \newline
\textcolor{comments}{{Unlike other low suprathermal electron flux cases, in cases 11 \& 12 there were significant emissions at 1027{\AA} (5 R \& 3 R), the source of which remains unclear. Electron impact on neutrals cannot be the source of the emissions as the brightness of other atomic lines, such as OI1304 and OI1356 would be significant as well. The low Lyman-$\alpha$/Lyman-$\beta$ ratios in these cases (2.6 - 3.3) rules out resonant scattering as a source (expected ratio of 300) and suggests photo-dissociation could be a driver of these emissions (expected ratio of 1.4). However, prompt-photodissociation would require significantly more \ce{H2O} along the line of sight (by a factor 5-15) to generate emissions in agreement with those observed. Late in the Rosetta mission, as these cases were, water was more weakly outgassing than \ce{CO2} in the southern hemisphere \citep{Gasc2017MNRAS, Laeuter2018}, so the required mixing ratios (up to 80\% water) are unlikely to have occurred and are inconsistent with the in situ measurements. Therefore, the source of these unexplained Ly-$\beta$ emissions remains an open question.}} \newline
\textcolor{comments}{{In the low electron flux cases (cases 9-13), the neutral column density is derived from a very simple model (Eq. \ref{eq: Haser column}), where we have assumed a fixed comet radius and constant neutral gas velocity. Using a constant expansion velocity may underestimate the gas density near the nucleus by up to 50\% \citep{Heritier2017IonComp}, but this has little impact on the conclusions in these low flux cases. We have also neglected any lateral motion of the neutral gas in the coma, which could impact the composition and density of the neutrals along the line of sight. When deriving column densities from in situ measurements, a more complete 3D structure of the neutral coma \citep[e.g. ][]{Bieler2015Comp3d}, including expansion of the gas near the nucleus, could be incorporated into the multi-instrument analysis.}} \newline
\textcolor{edits}{At large heliocentric distances}, the FUV spectra obtained in the southern hemisphere are very different to those from the northern, \textcolor{edits}{summer} hemisphere  due to the prominence of \ce{CO2} in the southern hemisphere \citep{Hassig2015}, especially post perihelion \citep{Gasc2017MNRAS}. Consequently, the FUV spectra contain much stronger atomic carbon lines (CI and CII) as well as molecular bands of CO, such as the Fourth Positive Group \citep{Feldman2018} compared to the northern hemisphere where they are barely detected. The hemispherical asymmetry in composition results in significantly different sources for several emission lines between the northern and southern hemispheres.\newline
OI1356 is produced primarily by \ce{e + CO2} in the southern hemisphere (see Fig.~\ref{fig: Nadir Quiet}a), whereas, in the northern hemisphere, \ce{e + O2} plays a more significant role \citep{Galand2020}. The OI1304 brightness had significant contributions from electron impact on all four of the major neutral species in the coma (\ce{CO2}, \ce{H2O}, \ce{CO} and \ce{O2}) across the selected cases over the shadowed nucleus in the southern hemisphere(see Fig.~\ref{fig: Nadir Quiet}b). The emissions near 1026~\si{\angstrom} correspond only to emissions of Lyman-$\beta$ from \ce{e + H2O} during the northern hemisphere \textcolor{edits}{summer} \citep{Galand2020}, whereas for post-perihelion cases in the southern, winter hemisphere, there can be a significant contribution of OI1027 from \ce{e + CO2} (see case 8, Fig.~\ref{tab: Nadir cases}c). Therefore, it is important to account for all four of the major neutral species when analysing FUV emission spectra over the southern hemisphere to derive column densities or mixing ratios. \newline
The CI1657 emissions are dominated by \ce{e + CO2}, when there is a large suprathermal electron flux (see Section \ref{sec: nadir 1-8}). However when the population of suprathermal electrons in the coma is small, fluorescence of \ce{CO} in the overlapping 4PG bands becomes a more significant driver of emissions near 1657~\si{\angstrom}. \newline
The brightness of the CII1335 emissions is highly sensitive to the column density of CO, due to the large emission frequency of \ce{CO} compared to \ce{CO2} in this line (see Fig.~\ref{fig: emission freq}e). The close agreement between the observed and modelled brightness of the CII1335 line confirms that the volume mixing ratio of CO derived from the ROSINA mass spectrometer is accurate, once the contribution to the \ce{CO} signal due to fragmentation of \ce{CO2} in the ion source is subtracted from the CO signal on the detector \citep{Dhooghe2014}. \newline
During the CIRs \textcolor{edits}{over the 9 and 10 July and on 4 August 2016}, the OI1356 and CI1657 emissions were dominated by \ce{e + CO2}, as attested by the agreement between the observed and modelled brightnesses (see Section \ref{sec: CIRs}). This is not surprising as \ce{CO2} was the dominant species outgassing from the southern hemisphere near the end of mission \citep{Luspay-Kuti2019}. For limb viewing, the Lyman-$\beta$ (Fig.~\ref{fig: August}c) and OI1304 (Fig.~\ref{fig: August}d) emissions both have significant contributions from electron impact, but the model underestimates the brightness of these lines. As the extended coma is also observed in limb viewing, resonant scattering of solar photons from atomic hydrogen and oxygen present along the line of sight may account for the discrepancy between the model and observations for these lines \citep{Combi2004, Feldman2018}. When viewing the illuminated nucleus, reflected solar flux from the nucleus introduced significant noise to the brightness of the CI1657 emissions, but the OI1356 line was generated only by dissociative excitation. \newline
Throughout the CIRs, the brightnesses of all the selected emission lines exhibit the same temporal variation as measured in the suprathermal electron flux (Section \ref{sec: CIRs}). \textcolor{comments}{{For all the time periods except the first on 4 Aug 2016, the modelled brightnesses agreed very closely with the observed line brightnesses, although there is a slight underestimation of OI1304 throughout the limb observations (see Figure \ref{fig: August}d), which may be attributed to resonant scattering from atomic oxygen in the extended coma. Resonant scattering off atomic hydrogen may contribute to Ly-$\beta$ emissions, but this is less apparent as the emissions are much weaker than those from the IPM, which have been accounted for. It is difficult to constrain resonant scattering off atomic oxygen or hydrogen as we do not have measurements of the density of these neutrals throughout the column. The discrepancy in magnitude in the first period in August may originate from some change in the electron flux throughout the column. There should be no significant degradation of electrons in the coma at the large heliocentric distances considered (see Section \ref{sec: electron flux}), but a large scale potential well \citep{Deca2017} could cause a variation of the electron flux over the extended column in limb viewing. However, if the assumption of a constant electron flux were invalid, the discrepancy should also arise in the other periods on 4 August, which have similar viewing geometries. }} \newline
In a limb or off-nadir viewing geometry, the emissions originate from a distant region of the coma, which cannot be probed in situ. The observed correlation between the in situ measurements of the electron flux and the remote measurements of the FUV spectrograph suggests that \textcolor{comments}{{any acceleration of the electron flux occurs on large scales as suggested by \cite{Deca2017}. At times with low electron fluxes, such as period 4 and the troughs in period 1 on 4 August 2016, there were no strong emissions of the FUV lines, which excludes any strong local heating of electrons in a distant region along the column.}} These results support the findings of \cite{Galand2020} that these are solar wind electrons, which undergo acceleration \textcolor{edits}{by several tens of \si{\electronvolt}} in the coma. Therefore, the FUV emissions are auroral in nature. The close correlation observed between the observed FUV auroral brightness and the electron flux allows FUV spectroscopy to be used as another measure of structures in the solar wind. \newline
The aurora observed in the southern hemisphere of 67P is similar to the diffuse aurora at Mars in several ways. Both auroras are driven by solar wind electrons on open draped field lines \citep{Brain2007, Volwerk2019} as shown in this study at comet 67P and by \cite{Schneider2015} at Mars. However, at comet 67P the solar wind electrons are accelerated by an ambipolar field \citep{Deca2017, Deca2019}, whereas at Mars the electrons driving diffuse auroras are accelerated at the Sun rather than within the Martian system \citep{Schneider2015}. \textcolor{comments}{{Consequently, these auroras are persistent at comet 67P whereas at Mars the diffuse aurora is only observed during strong solar events.}}

\section*{Acknowledgements}
Rosetta is a European Space Agency (ESA) mission with contributions from its member states and the National Aeronautics and Space Administration (NASA).We acknowledge the continuous support of the Rosetta teams at the European Space Operations Centre in Darmstadt and at the European Space Astronomy Centre. We acknowledge the staff of CDDP and Imperial College for the use of AMDA and the RPC Quicklook database. Work at Imperial College London was supported by STFC of UK under grant ST/N000692/1 and ST/S505432/1. The Alice team acknowledges support from NASA’s Jet Propulsion Laboratory through contract 1336850 to the Southwest Research Institute. AB was supported by the Swedish National Space Agency (grant 108/18). MR acknowledges the support of the State of Bern and the Swiss National Science Foundation (200021 165869, 200020 182418). VIRTIS was built by a consortium, which includes Italy, France, and Germany, under the scientific responsibility of the Istituto di Astrofisica e Planetologia Spaziali of INAF, Italy, which also guides the scientific operations. The VIRTIS instrument development, led by the prime contractor Leonardo-Finmeccanica (Florence, Italy), has been funded and managed by ASI, with contributions from Observatoire de Meudon financed by CNES, and from DLR. We thank the Rosetta Science Ground Segment and the Rosetta Mission Operations Centre for their support throughout all the phases of the mission. The VIRTIS calibrated data will be available through the ESAs Planetary Science Archive (PSA) Website (www.rssd.esa.int) and is available upon request until posted to the archive. We thank the following institutions and agencies for support of this work: Italian Space Agency (ASI, Italy) contract number I/024/12/1, Centre National d’Etudes Spatiales (CNES, France), DLR (Germany), NASA (USA) Rosetta Program, and Science and Technology Facilities Council (UK). All ROSINA data are the work of the international ROSINA team (scientists, engineers and technicians from Switzerland, France, Germany, Belgium and the US) over the past 25 years, which we herewith gratefully acknowledge.
\section*{Supporting data}
All of the primary data from Rosetta used in this study will be available through ESA's Planetary Science Archive (PSA) at \url{www.rssd.esa.int} or is available on request until posted to the archive. The processed emission line brightnesses, emission frequencies, column densities, and electron particle fluxes are made available through the Centre de Donn\'{e}s astronomiques de Strasbourg (CDS) archives. 
\bibliography{my_collection}

 \begin{appendix} 
 \section{Calculating the suprathermal electron flux from the RPC/IES counts}\label{sec: L2 to L3}

The Rosetta Plasma Consortium Ion and Electron Sensor (RPC/IES) was a top hat electrostatic analyser which measured the number of electron counts in bins of azimuthal angle, elevation angle and energy \citep{Burch2007}. The sensor comprised $16\times16\times128$ bins in these dimensions with an angular resolution of $\ang{22.5}\times\ang{5}$. The energy bins were distributed approximately logarithmically with a bin width $\Delta E/E =8\%$. RPC/IES measured the electron count rate in all 16 azimuthal bins simultaneously, whilst the electron and energy bins were cycled through. During the mission, anodes from the 8$^\text{th}$ to 15$^\text{th}$ azimuthal bins degraded \citep{Broiles2016a}, hence we only used anodes 0 to 7 in our analysis. \newline
The electron particle flux was derived from RPC/IES measurements by \cite{Heritier2018}, but here we provide an outline of each step in this calculation.
Within the Level 3 files on the PSA, the background count of electrons, $C_{L3BG}$, is given for each scan of the electron spectrometer, which is independent of the instrument bin. The background count must be corrected as below before being subtracted from the counts in the measured bins.
\begin{equation}\label{eq: BG count}
        C_{BG} = C_{L3BG} + \sqrt{5}C_{L3BG}
\end{equation}
When installed on Rosetta, the field of view of the instrument was obscured by the main body of the spacecraft and other instruments \citep{Clark2015}. This restricted field of view has been corrected with geometric factors, $G(\theta_0, \phi_0, E_0)$ \citep{Broiles2016a}, where the subscript $0$ indicates that this has been evaluated on the underlying instrument grid of $16\times16\times128$. However, in many of the operation modes, neighbouring bins in all three dimensions were collapsed into one datapoint to reduce the demand on telemetry. Combined azimuthal bins have already been separated in the Level 2 datafiles, available on the PSA but the combined bins in elevation and energy have not been separated. We refer to $N_{bins}$, the number of combined energy bins multiplied by the number of combined elevation bins. \newline
The counts, with background removed on the binned grid, were then un-binned onto the instrumental $16\times16\times128$ grid, assuming the electrons were distributed evenly between the combined bins. 
\begin{equation}
    C'(\theta_0,\phi_0, E_0, t) = \frac{C(\theta_{b},\phi_{b}, E_{b}, t) - C_{BG}(t)}{N_{Bins}},
\end{equation}
where the subscript $b$ refers to the binned grid. The un-binned counts were then divided by the geometric factor for each bin and the integration time associated with the operation mode to convert to a differential energy flux, $g'(\theta_0, \phi_0, E_0, t)$. The geometric factor of the instrument used here is based on observations on 1 January 2015, as derived by \cite{Broiles2016a}.
\begin{equation}
    g'(\theta_0, \phi_0, E_0, t) = \frac{ C'(\theta_0,\phi_0, E_0, t)}{\Delta t \times G(\theta_0, \phi_0, E_0)}
\end{equation}
The differential flux above has not accounted for the efficiency, $\epsilon(E)$, of the MicroChannel Plates (MCPs) in the instrument, which is dependent on the impacting electron energy, E \citep[in \si{\electronvolt};][]{Kurz1979}.
\begin{equation}\label{eq: MCP efficiency}
\begin{split}
\epsilon(E)=0.52 & \exp\Big( -[\log(E+37)-2.3]^2 / 0.82 \Big) + 0.071 \\ & + 0.16\log(E+37)-0.028\log(E+37)^2
\end{split}
\end{equation}
We returned to the binned energy grid, $E_{b}$, by summing over combined energy bins, before dividing by the efficiency to give the differential electron particle flux (DEF), $j(\theta_b,\phi_b, E_b)$.
\begin{equation}
j(\theta_0, \phi_0, E_b, t) = \frac{ \sum\limits_{\text{Comb. Energies}} g'(\theta_0,\phi_0, E_0, t)}{N_{\text{Comb. Energies}}\times \epsilon(\bar{E}_b) \times \bar{E}_b},
\end{equation}
where the summation is over the energy bins which were combined in the binned grid. $\bar{E_b}$ is the average energy of the combined energy bins.
The DEF was integrated over the field of view of RPC/IES to give the electron particle flux, $J(E)$ \citep[in \si{\per\square\centi\metre \per\second \per\electronvolt};][]{Heritier2018}:
\begin{equation}\label{eq: EPlFlux}
J(E_b) =\frac{4\pi}{2\pi\sin\ang{48}}\int_{\theta_0 = \ang{-48}}^{\ang{48}}\int_{\phi_0 = \ang{0}}^{\ang{180}}\!\cos\theta_0\times j(\theta_0, \phi_0, E_b)\, \mathrm{d}\phi_0\mathrm{d}\theta_0
\end{equation}
The term representing a geometric factor at the start of Eq. \ref{eq: EPlFlux} is a correction for the limited field of view of RPC/IES, under the assumption of an isotropic electron particle flux at the detector.\newline
We note that the electron particle flux at the detector was not representative of the electron flux in the coma, as the Rosetta spacecraft was generally at a spacecraft potential lower than -10\,\si{\volt} \citep{Odelstad2015b}. This was corrected for using Liouville's theorem (see Eq. \ref{eq: Liouville}) but this prevented measurement of any electrons at low energies (see Section \ref{sec: electron flux}). 
\section{Removing solar contamination from FUV emission spectra}\label{sec: Alice analysis}
\begin{figure}[htbp]
    \includegraphics[width = 0.5 \textwidth]{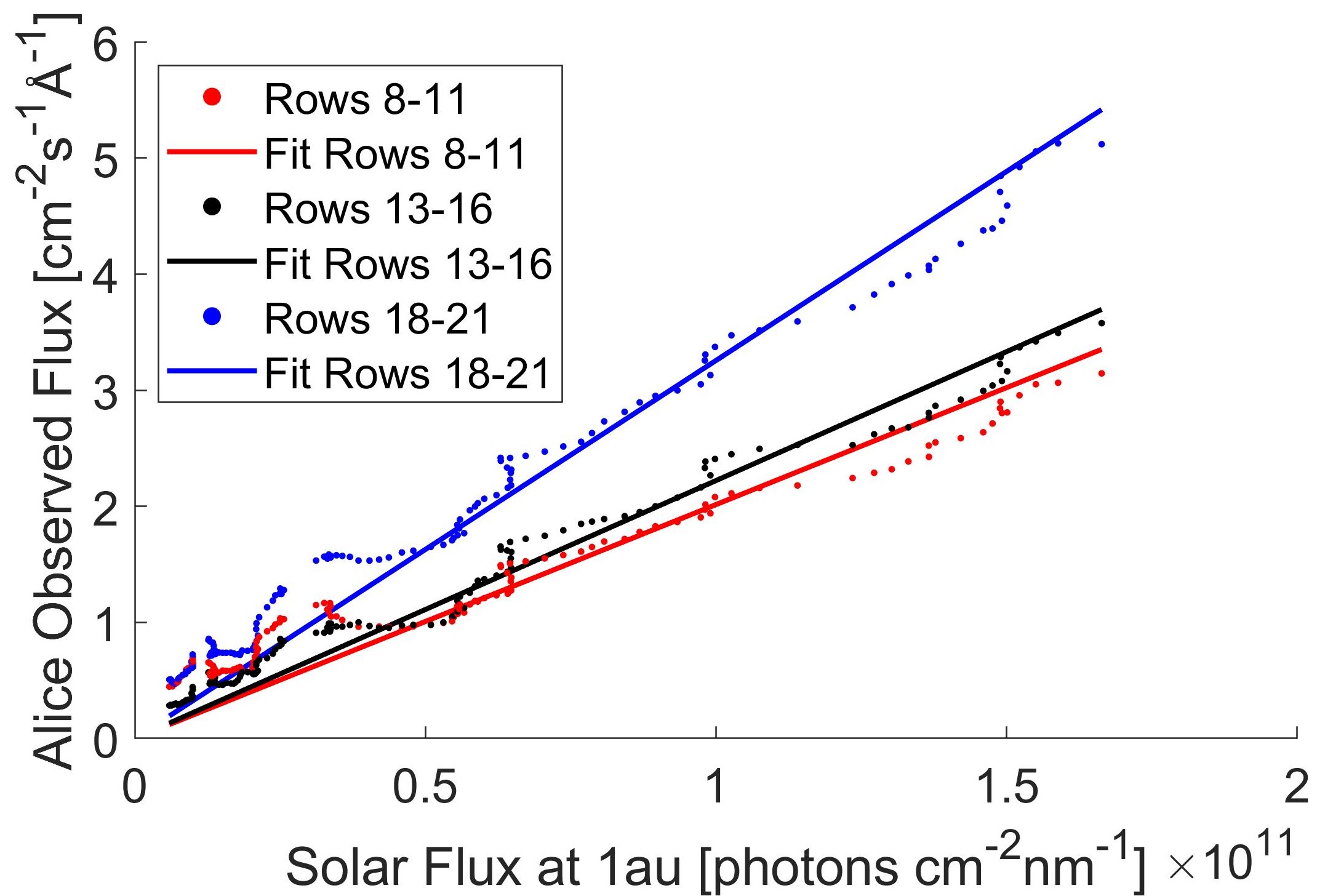}
    \caption{\textcolor{comments}{{Comparison of the photon flux observed by Alice at 15:58~UT 9 July 2016 to the solar photon flux, measured by TIMED-SEE at long wavelengths (1500{\AA} - 1800 {\AA}). Three distinct regions of the Alice slit (Rows 8-11 [red], Rows 13-16 [black] and Rows 18-21 [blue]) have been plotted which are the same as those plotted in Figure \ref{fig: July}. The linear fits are used to scale the solar flux, so it can be subtracted from the observed Alice spectrum.}}}
    \label{fig: solar flux fits}
\end{figure}
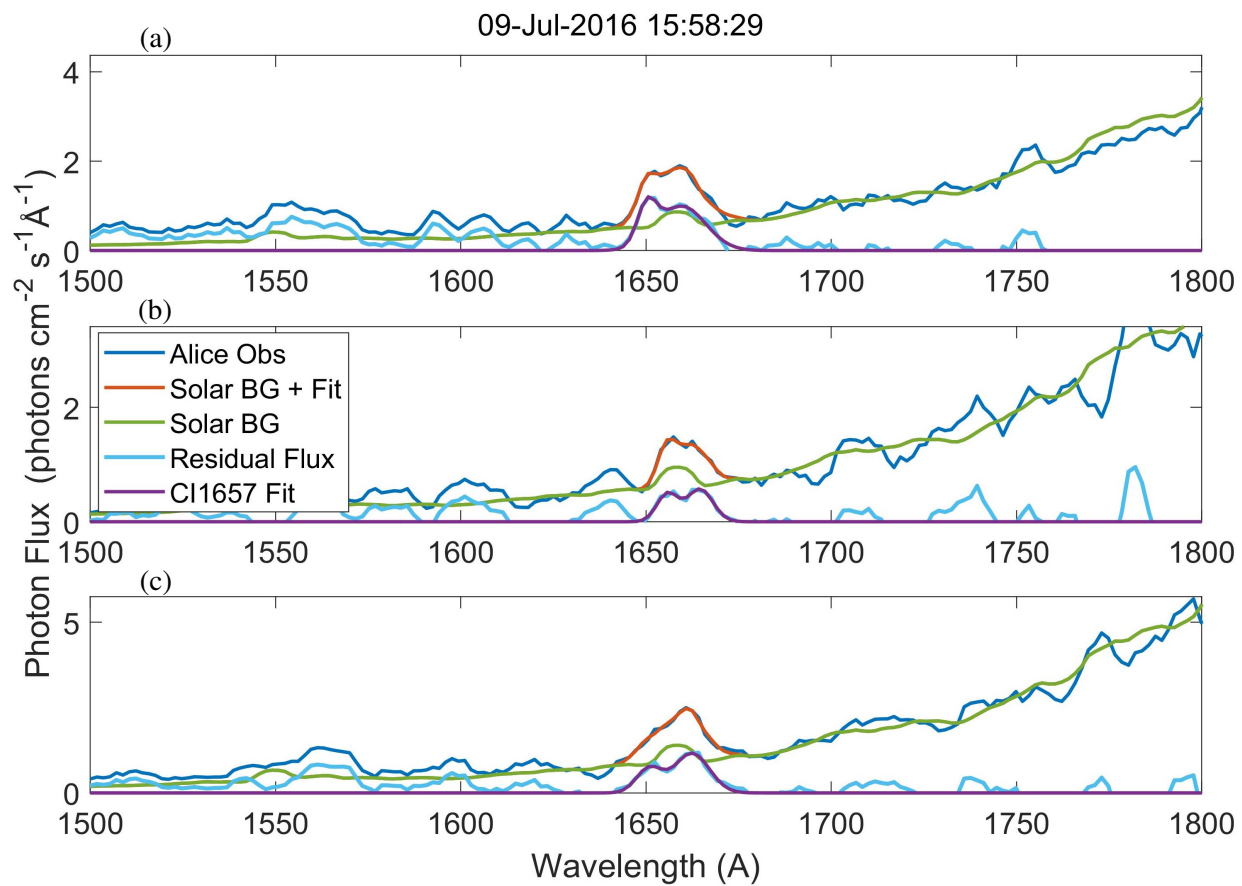
\begin{figure*}[htbp]
    \centering
    \begin{tikzpicture}[node distance= 1cm, baseline=(current  bounding  box.center)]
    \node (jul) at (0,0) {\includegraphics[width = 0.9\textwidth]{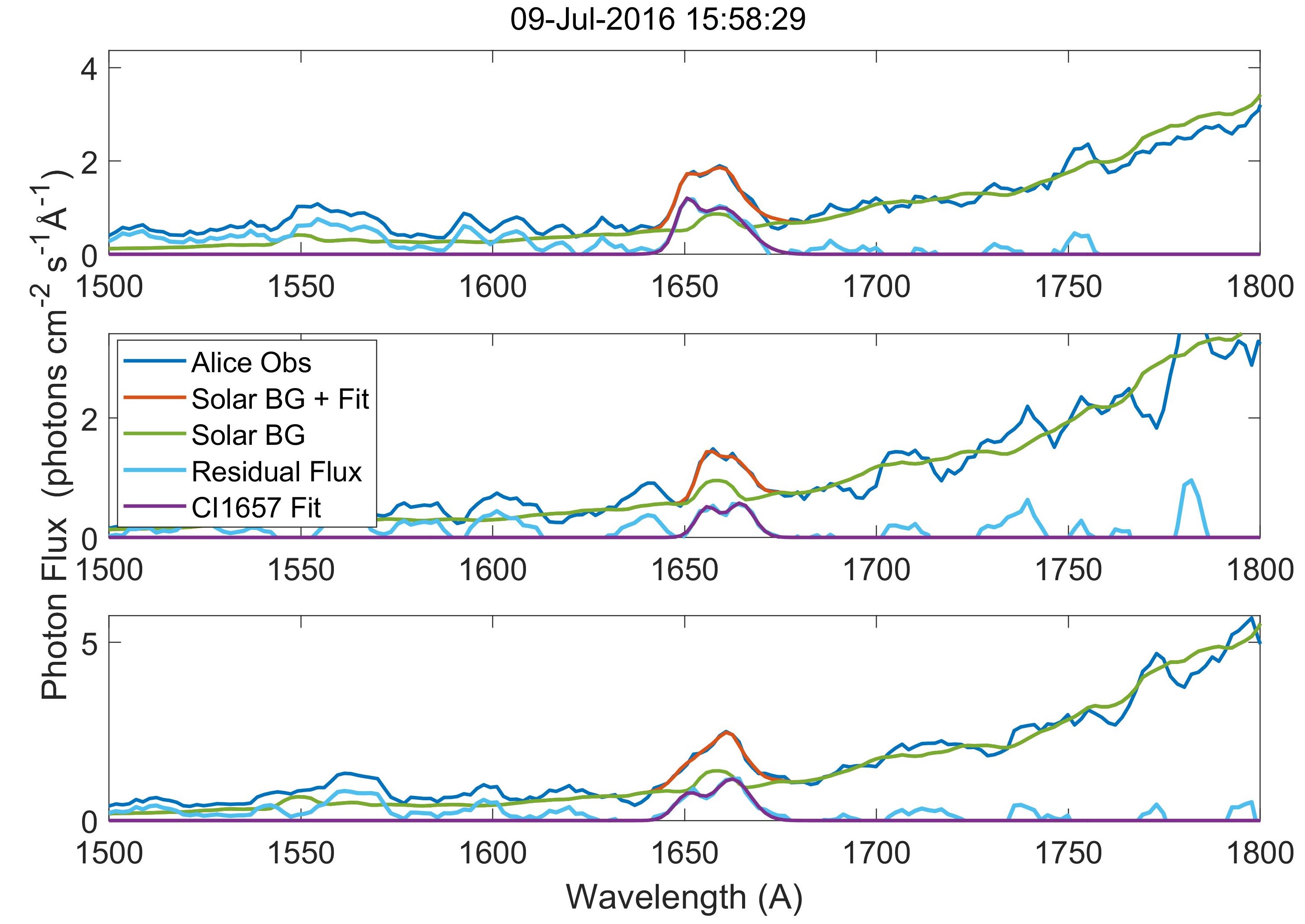}};
    \node [above left = 11 and 5.7 of jul.base, font =\large] (A) {(a)};
    \node [above left = 7.4 and 5.7 of jul.base, font =\large] (B) {(b)};
    \node [above left = 3.8 and 5.7 of jul.base, font =\large] (C) {(c)};
    \end{tikzpicture}
    \caption{\textcolor{comments}{{Breakdown of the fitting procedure to retrieve the CI1657 and CO4PG line brightness for (a) Rows 8-11, (b) Rows 13-16 and (c) Rows 18-21. The solar spectrum from TIMED-SEE (green) is smoothed and then scaled to the Alice photon flux (dark blue) at long wavelengths. The residual observed flux, once the solar spectrum is subtracted, is shown in light blue and with the CI1657 fit plotted in purple. The combination of the solar background with the CI1657 line fit is plotted in orange.}}}
    \label{fig: July CI1657}
\end{figure*}

\textcolor{comments}{{The emission spectra gathered on 9-10 July 2016 were strongly polluted by reflected solar flux from the illuminated surface of the nucleus. As we are only interested in the emissions from the coma, it was necessary to remove any contribution of reflected solar photons. The shape of the solar spectrum was taken from TIMED-SEE measurements on 16 July 2016 (green, Figure \ref{fig: July CI1657}), which was at the same Carrington longitude as the Alice observations. As this study considers only cases at large heliocentric distances, the coma was optically thin and there should not have been significant absorption of the solar flux by cometary neutrals \citep{Heritier2018a}.}} \newline
\textcolor{comments}{{As the solar flux was more intense at the long wavelength end of the Alice spectral range \citep{Feldman2018}, the photon flux at 1au was compared to the FUV emission spectra from Alice between 1500 {\AA} and 1800 {\AA} for each set of co-added rows (see Figure. \ref{fig: solar flux fits}), with bright emission features around 1561 {\AA} and 1657 {\AA} excluded. A clear linear correlation was seen for all three regions of the Alice slit illustrating the strong presence of the solar spectrum. The reflected solar flux was derived from the linear fit (green, Fig. \ref{fig: July CI1657}) and was subtracted from the observed Alice spectra. The residual FUV emission spectra (light blue, Fig. \ref{fig: July CI1657}) were then fitted with a linear combination of two gaussian distributions (purple). \newline
The CI1657 emission, measured by Alice (dark blue, Fig. \ref{fig: July CI1657}), was more diffuse than the atomic line measured at 1au by TIMED-SEE as a result of the lower spectral resolution (8 -12 {\AA}). The emission feature in the Alice spectra may also contain several bands of the CO Fourth Positive Group (see Section \ref{sec: Alice Obs}) leading to further broadening. We smoothed the TIMED-SEE flux over 10~{\AA} to broaden the peak at 1657~{\AA}, which better captured the width of the peak in the Alice dataset.
If the adjusted goodness of fit exceeded 0.65, the line brightness is given by 
\begin{equation}
    B^X = (A_1 + A_2) \times 4 \arcsin\Big[\sin\Big(\frac{\alpha}{2}\Big)\sin\Big(\frac{\beta}{2}\Big)\Big] \times \frac{10^6}{4\pi}
\end{equation}
where the line brightness $B^X$ is in rayleighs and the amplitudes of the Gaussian distributions, $A_1$ and $A_2$, have units of photons~cm$^{-2}$s$^{-1}$. $\alpha$ and $\beta$ are the angles subtended by the coadded rows in the Alice slit, and the numerical factor originates from the definition of 1 rayleigh (see Equation \ref{eq: model brightness})}}.


 \end{appendix}
\end{document}